\documentclass[11pt,hyper]{JHEP3}
\usepackage{amsmath,epsfig}
\usepackage{amssymb,amsfonts}
\usepackage{graphicx}
\usepackage{multirow,cite}
\usepackage{subfigure}
\usepackage{datetime}

\newcommand{\bi}{\begin{itemize}}
\newcommand{\ei}{\end{itemize}}
\newcommand{\non}{\nonumber}
\def\p{\partial}
\def\a{\alpha}
\def\b{\beta}
\def\d{\delta}
\def\g{\gamma}
\def\l{\lambda}
\def\e{\epsilon}

\def\th{\theta}
\def\om{\omega}
\def\s{\sigma}
\def\A{\mathcal{A}}
\def\B{\mathcal{B}}

\def\V{\mathcal{V}}
\def\M{\mathcal{M}}

\def\slr{SL(2,\mathbb{R})}

\def\S{\Sigma}
\def\Om{\Omega}
\def\r{\rightarrow}
\def\half{{\frac12}}

\newcommand{\N}{{\mathcal{N}}}

\newcommand{\bea}{\begin{eqnarray}}
\newcommand{\eea}{\end{eqnarray}}
\newcommand{\be}{\begin{equation}}
\newcommand{\ee}{\end{equation}}

\def\ds{\displaystyle}
\def\IR{\mathbb{R}}

\title{Un-twisting the NHEK with spectral flows}
\preprint{IPhT - T12/022}
\author{Iosif Bena$^\dag$, Monica Guica$^\flat$ and Wei Song$^\ddag$
\vspace{0.5cm}

{\it $^\dag$ Institut de Physique Th\'eorique,  CEA Saclay, CNRS-URA 2306,  \vspace{-2 mm}\\
 \hspace{-0.25 cm} 91191 Gif sur Yvette, France
 }

 \vspace{ 3 mm}

{\it $^\flat$ Department of Physics and Astronomy, University of Pennsylvania,   \vspace{-2 mm} \\
\hspace{-0.25 cm} Philadelphia, PA 19104-6396, USA
}

 \vspace{ 3 mm}

{\it $^\ddag$ Center for the Fundamental Laws of Nature, Harvard University,  \vspace{-2 mm} \\
 \hspace{-0.25 cm} Cambridge, MA 02138, USA
} }

\abstract{

\vskip 5 mm

We show that the six-dimensional uplift of the five-dimensional Near-Horizon-Extremal-Kerr (NHEK) spacetime can be obtained from an $AdS_3 \times S^3$ solution by a sequence of supergravity - but not string theory - dualities. We present three ways of viewing these pseudo-dualities: as a series of transformations in the STU  model, as a combination of Melvin twists and T-dualities and, finally, as a sequence of two generalized spectral flows and a coordinate transformation. We then use these to find an infinite family of asymptotically flat embeddings of NHEK spacetimes in string theory, parameterized by the arbitrary values of the moduli at infinity. Our construction reveals the existence of non-perturbative deformations of asymptotically-NHEK spacetimes, which correspond to the bubbling of nontrivial cycles wrapped by flux, and paves the way for finding a microscopic field theory dual to NHEK which involves Melvin twists of the D1-D5 gauge theory. Our analysis also clarifies the meaning of the generalized spectral flow solution-generating techniques that have been recently employed  in the literature.}

\begin{document}{

\section{Introduction and summary}

One of the great successes of string theory has been to provide a microscopic explanation for the Bekenstein-Hawking entropy of black holes \cite{microorigin}. Nevertheless, most of the black holes that have been understood so far have a near-horizon geometry that contains an $AdS_3$ factor (in some duality frame), and hence their microscopic description is in terms of a dual two-dimensional CFT  \cite{Strominger:1997eq,Maldacena:1997re}.

More recently, it has been conjectured that general extremal black holes, whose near horizon geometry does not contain an $AdS_3$ factor, are also described by a dual CFT \cite{kerrcft} (see also \cite{Hartman:2008pb,Lu:2008jk,Azeyanagi:2008kb,Chow:2008dp,Azeyanagi:2008dk, Isono:2008kx,Chen:2009xja,Lu:2009gj,Compere:2009dp,Krishnan:2009tj,Astefanesei:2009sh,Wen:2009qc,Azeyanagi:2009wf,Peng:2009wx}).
This conjecture applies in particular to four-dimensional extremal and near-extremal Kerr black holes, astrophysical examples of which have already been observed in our galaxy. The near-horizon geometry of general extremal black holes contains  a {\it warped} $AdS_3$ factor - that is, a  space which is a $U(1)$ fiber over $AdS_2$ \cite{bardhor,reallkund} - which is believed to play a key role in the conjectured holographic duality.

The original conjecture of \cite{kerrcft} was based on a careful analysis of the asymptotic symmetries of the near-horizon region of the extreme Kerr black hole (NHEK) \cite{bardhor}. These symmetries were shown to generate a Virasoro algebra, whose central charge correctly reproduces the Bekenstein-Hawking entropy of the black hole \emph{via} Cardy's formula. Assuming in addition unitarity and locality of the dual field theory, it seemed rather obvious that it should be a conformal field theory.

Nevertheless, the existence of a spacetime Virasoro symmetry and the applicability of Cardy's formula does not always imply the existence of a dual holographic CFT of the usual type, as exemplified by  ``dipole CFTs'' \cite{stromdipwei}. These theories are the IR limit of some rather exotic nonlocal field theories called dipole theories \cite{berggan}, which are related by  a ``pseudo-duality'' known as TsT\footnote{If the shift were an integer the TsT would be a true duality of string theory, but for dipole theories this shift is not integer. However, pseudo-dualities are still symmetries of the classical supergravity action, and in particular can be used to take solutions into solutions.} (T-duality, shift, T-duality back) to the D1-D5 gauge theory. One peculiarity  of dipole CFTs is that operators have momentum-dependent conformal dimensions, which indicates that they are not usual CFTs, but rather they resemble non-relativistic ones \cite{kerrdip}. Yet, dipole CFTs are dual to a warped $AdS_3$ spacetime which admits a Virasoro asymptotic symmetry group \cite{stromdipwei,kerrdip} that correctly reproduces the black hole entropy.

Hence, in order to  understand the kind of ``CFT'' dual to a Kerr black hole it is very useful to have a concrete example, if not of the ``CFT'', then at least of a theory that flows to it in the infrared. In this article we take two steps in this direction: first, we construct large classes of brane configurations that have a Kerr near-horizon geometry; and second, we  argue that the theory obtained by performing certain Melvin twists on the D1-D5 gauge theory - which are known to generate  certain fermion mass deformations among other things - flows in the infrared to the theory dual to a Kerr black hole.

There are many examples of extremal black holes that fall within the scope of the Kerr-CFT proposal, and here we will focus on a particular one: the six-dimensional uplift of the non-supersymmetric extremal Kerr-Newman black hole in five dimensions with nonzero $J_L$ but $J_R =0$.  The near-horizon geometry of this black hole (which we will still call NHEK or, to avoid confusion, $6d$ NHEK) is simpler and more symmetric than that of the four-dimensional Kerr black hole, mainly because the warped $AdS_3$ factor it contains has a constant, rather than polar-angle-dependent, warp factor. This black hole can be embedded in string theory as an extremal non-supersymmetric D1-D5-p  black hole with the above angular momenta, and constitutes a natural departure point for concretely understanding the microscopic theory dual to Kerr \cite{microkerr}.

The key observation of this paper is that the near-horizon of the six-dimensional uplift of the five-dimensional Kerr-Newman black hole can be related via a set of string pseudo-dualities to $AdS_3 \times S^3$, which is the near horizon of the D1-D5 system.

There are three different ways of viewing these pseudo-dualities: The first is to view both $AdS_3 \times S^3$ and $6d$ NHEK as $T^2$ fibrations over an $AdS_2 \times S^2$ base. Reducing along the $T^2$ fiber, they become solutions of the four-dimensional STU model. These solutions are related by the $SL(2,\mathbb{R})^3$ symmetry of the STU model, but cannot be related by just $SL(2,\mathbb{Z})^3$ dualities. Consequently, $6d$ NHEK is related by pseudo-dualities to $AdS_3 \times S^3$. These pseudo-dualities are a combination of S-dualities, coordinate changes, TsT transformations and six-dimensional electromagnetic dualities, whose microscopic interpretation in the D1-D5-p system is far from obvious.

Nevertheless, we  show that the combination of S-duality, TsT and S-duality (which appears twice in the duality chain above) applied to the D1-D5-p system is in fact the same as a T-duality along the common D1-D5 direction, M-theory uplift, compactification back to type IIA string theory with a Melvin twist, followed by a T-duality back\footnote{This identity of dualities had already been observed for certain D-brane probes in \cite{ganpuff,aspuff}.}. Applying this T-Melvin-T transformation once yields a warped $AdS$ geometry that is not the $6d$ NHEK geometry. To obtain $6d$ NHEK one needs to repeat this procedure, but this time T-dualize both along the common D1-D5  direction and along the  $T^4$ or  $K^3$ wrapped by the D5 branes, uplift to M-theory, re-descend with a Melvin twist, and T-dualize five times back.

Yet another way of understanding the above pseudo-dualities is to realize that they precisely correspond to generalized spectral flows, of the type introduced in \cite{Bena:2008wt} and originally used to relate supersymmetric multi-center solutions. The advantage of viewing the dualities as  generalized spectral flows is that one understands very well the action of the latter on the D-branes underlying a given solution. Hence, one can use the fact that the $AdS_3 \times S^3$ geometry is supported by  D1 and D5 branes and momentum and can be deformed by moving some of these branes away, to infer the types of branes that give rise to the $6d$ NHEK geometry and to find some of its deformations.

These ways of understanding the duality allow us to extract three interesting new pieces of physics. First, it is known that, unlike the usual spectral flows, generalized spectral flows do not preserve the asymptotically $AdS_3 \times S^3$ structure of solutions. They do preserve, however, the asymptotically $\IR^{3,1} \times U(1)^2$ structure of spacetimes \cite{Bena:2008wt}. Hence, if we embed the original D1-D5-p system in Taub-NUT - whose asymptotic structure will not change - and then perform the spectral flows, we obtain a family of asymptotically-flat solutions with a NHEK infrared, parameterized by the arbitrary values for the moduli at infinity. 
These solutions generalize the  brane configuration with a NHEK near-horizon found in \cite{stromwei}.  Their structure  is rather nontrivial, and they are T-dual to the general class  of solutions that have been obtained in \cite{DGR}. The advantage of having a general solution is that it is easier to explore and interpret its various limits, in which the dual brane interpretation may simplify. We plan to address this issue in future work.

The second feature of the NHEK geometries that we explore is their space of deformations. Indeed, it is well known that the $AdS_3 \times S^3$ solution can be deformed by taking for example some of the D1 or D5 branes and moving them on the Coulomb branch. This deformation takes one from the maximally twisted sector of the CFT (that has the highest entropy in the Cardy regime) to a different sector of the same CFT. It is very easy to use the generalized spectral flows to map certain Coulomb branch configurations of the original solution to certain configurations in the moduli space of deformations of the NHEK geometry. These configurations have a different topology than the original geometry, and have nontrivial ``bubbles''  wrapped by flux in the infrared. Note that these deformations are not visible in a purely perturbative analysis of the $6d$ NHEK geometry, and thus evade the ``no dynamics'' constraint discussed in \cite{Amsel:2009ev,Dias:2009ex}.

The third use of the duality chain we found is to indicate a way to construct a field theory that flows in the infrared to the CFT dual to extremal five-dimensional Kerr-Newman black holes. To see how this may work, remember that if one takes a stack of D4 branes, uplifts them to M-theory and reduces back with a Melvin twist, one obtains the supergravity dual of the mass-deformed D4 brane theory, which is more precisely $\N=1$ Super Yang Mills in five dimensions with a massive chiral multiplet \cite{chocky}.  If one used  a similar analysis to find the deformations corresponding to the first and the  second Melvin twist in the field theory describing the D1-D5 system, one would obtain a theory that flows in the infrared to the CFT dual to NHEK, much like the undeformed D1-D5 field theory flows in the infrared to the D1-D5 CFT. We plan to expand our investigation of this in the future.

Finally, we should mention that the first generalized spectral flow has a simple and interesting S-dual interpretation in terms of  the  $SL(2,\mathbb{R}) \times SU(2)$ WZW model that lives on the worldsheet of a string propagating in the $AdS_3 \times S^3$ background supported by NS-NS flux. The effect of the TsT transformation in this context  is understood,  and it can be shown to yield a worldsheet CFT which is related to the original WZW model by a redefinition of the worldsheet coordinates that does not respect the periodicities of the angular coordinates \cite{Frolov:2005dj,alday}. It follows that string theory in this warped $AdS_3$ background  is equivalent to the string theory propagating in $AdS_3 \times S^3$ with non-local fields. It would be very interesting to better understand the physical consequences of this observation \cite{weiwants}.

As a byproduct of our analysis, we also obtain a simpler definition of generalized spectral flows in terms of string (pseudo)-dualities. This allows us to write the effect of generalized spectral flows on any background with two commuting isometries in closed form, up to solving a small set of linear differential equations. This approach is much simpler and more intuitive than those previously used in \cite{Bena:2008wt,DGR}, and can be applied to any black hole geometries, including non-extremal ones. 

\bigskip

This paper is structured as follows: in section \ref{adstonhek}, we review the $SL(2,\mathbb{R}) \times SU(2)$ invariant solutions of type IIB supergravity of \cite{kerrdip} and show that they are all related by STU transformations. In particular, we find the explicit STU matrices  which relate a given $6d$ NHEK geometry to $AdS_3 \times S^3$. In section \ref{spflmel}, we discuss generalized spectral flows and relate them to STsTS and T-duality - Melvin twist - T-duality transformations. We also develop a general formula for the action of generalized spectral flows on any six-dimensional geometry with two commuting isometries. In section \ref{spfld1d5} we exemplify our technique by finding the effect of two generalized spectral flows on the D1-D5-p-KK metric and we compare our answer with previous results in the literature in section \ref{infam}. Also in section \ref{infam}, we present a generalization of the spectral-flowed D1-D5-p-KK metric which has arbitrary values of the axions at infinity. This gives the largest known family of NHEK embeddings in string theory. Finally, in section \ref{micfeat}, we use our technology to analyze some microscopic features of $6d$ NHEK spacetimes, in particular to find a branch of their moduli space of deformations which consists in the bubbling of nontrivial cycles wrapped by flux. Various details of the calculations are collected in the appendices.

\bigskip

\section{From  $AdS_3 \times S^3$  to NHEK via STU transformations \label{adstonhek}}

In this section we review the $SL(2,\mathbb{R}) \times SU(2)$ invariant solutions of type IIB supergravity presented in \cite{kerrdip} and show all such backgrounds are related by string
pseudo-dualities, and more precisely by STU transformations.

\subsection{Warped  backgrounds of type IIB with NS flux \label{eom}}

\subsubsection*{Preliminaries}

We consider solutions of type IIB supergravity compactified on a four-dimensional manifold $M_4$, where $M_4$ can be either $T^4 $ or K3. Moreover, we consider a consistent truncation  of the type IIB supergravity action to six dimensions, which contains only NS-NS sector fields. The six-dimensional action is\footnote{The six-dimensional  dilaton $\Phi = \half(\phi_1 + \phi_2)$, where $\phi_1$ is the ten-dimensional dilaton and $e^{-\phi_2}$ is the Einstein frame volume of $M_4$.} \cite{Duff:1995sm}

\be
S = \int d^6 x \sqrt{-g} \,e^{-2 \Phi} \,\left( R + 4 (\p \Phi)^2 - \frac{1}{12} H^2 \right) \label{6dact}
\ee
The above action has a $Z_2$ duality symmetry, which acts as \cite{Duff:1998cr}

\be
\Phi \r - \Phi \;, \;\;\;\;\; H \r e^{-2\Phi} \star H\;, \;\;\;\;\;
g_{\mu\nu} \r e^{-2\Phi} g_{\mu\nu} \label{sduality}
\ee
A simple solution to the equations of motion is $AdS_3 \times S^3$, of radius $2 \ell$

\be
ds^2 = \ell^2 \bigl( - w_+ w_- +  w_3^2 + \s_1^2 + \s_2^2 + \s_3^2 \bigr) \;, \;\;\;\; \non
 \ee

\be
H = \ell^2 \bigl( \s_1 \wedge \s_2 \wedge \s_3  - \half w_+ \wedge w_- \wedge w_3\bigr) \;, \;\;\;\;\; H = \star H \label{ads3s3}
\ee
The six-dimensional dilaton $\Phi$ is attracted, but is not determined\footnote{The $AdS_3\times S^3$ solution appears in the near horizon of the NS1-NS5 system. There, $e^{2\Phi} = \frac{Q_5}{Q_1}$, whereas $\ell^2= Q_5$.} by $\ell$. In the above equations,  the $w_i$ are left-invariant (i.e. $SL(2,\mathbb{R})_L$-invariant ) one-forms on $AdS_3$

\be
w_+ = - e^{- y} \left(\frac{d\rho}{\rho} + \rho\, d\tau \right) \;, \;\;\;\;\; w_- =e^{ y} \left(\frac{d\rho}{\rho} - \rho \, d\tau \right) \;, \;\;\;\;\; w_3 = d y + \rho \,d\tau
\ee
and the $\s_i$ are left-invariant ($SU(2)_L$-invariant) one-forms on $S^3$

\be
\s_1 = \cos \psi \, d\th + \sin \th \sin \psi\, d\phi \;, \;\;\;\; \s_2 = -\sin \psi \, d\th + \sin \th \cos \psi\, d\phi \non
\ee

\be \s_3 = d\psi + \cos \th \, d \phi \label{su2l}
\ee
The same solution can be rewritten in terms of right-invariant ($SU(2)_R$-invariant) one-forms on $S^3$

\be
\s_1' = \cos \phi \, d\th + \sin \th \sin \phi\, d\psi \;, \;\;\;\; \s_2' = -\sin \phi \, d\th + \sin \th \cos \phi\, d\psi \non
\ee

\be
 \s_3' = d\phi + \cos \th \, d \psi \label{su2r}
\ee
which are obtained from the $\s_i$ by interchanging $\psi$ and $\phi$. The metric has the same expression as before, with $\s_i \r \s_i'$ . Nevertheless,  given that

\be
\s_1 \wedge \s_2 \wedge \s_3 = - \s_1' \wedge \s_2' \wedge \s_3'
\ee
the solution for the field $H$ is now\footnote{For the purposes of this section, we use the convention $\e_{\tau \rho  y \th \phi \psi}=1$.}

\be
H = \ell^2 \bigl(-\, \s_1' \wedge \s_2' \wedge \s_3'  - \half w_+ \wedge w_- \wedge w_3\bigr)
\ee
Thus, $AdS_3 \times S^3$ can be written both in a manifestly-$SU(2)_L$ or in a manifestly-$SU(2)_R$ invariant way. As a shortcut, we can encompass both expressions in one formula by writing

\be
H = \ell^2 \bigl(\varepsilon \, \s_1 \wedge \s_2 \wedge \s_3  - \half w_+ \wedge w_- \wedge w_3\bigr)
\ee
with the understanding that when $\varepsilon =1$, the $\s_i$ stand for the $SU(2)_L$-invariant one-forms on $S^3$ \eqref{su2l}, while when $\varepsilon =-1$, they stand for the $SU(2)_R$-invariant one-forms $\s_i'$ in \eqref{su2r}. This notation allows us to write the solutions with both manifest $\slr_L \times SU(2)_R$ and   $\slr_L \times SU(2)_L$ invariance in a compact and unified manner.

Our conventions for left and right are such that eight of the Killing spinors of $AdS_3 \times S^3 \times M_4$ are $SL(2,\mathbb{R})_L \times SU(2)_L$ invariant, whereas the other eight are $SL(2,\mathbb{R})_R \times SU(2)_R$ invariant. In the following, we will be mostly interested in quotients of $AdS_3$ and its warped generalizations by an $\slr_R$ element of the isometry group, whose effect is to make $y$ compact and break the right-moving supersymmetries. We will also be considering more general quotients, which simultaneously act on the Hopf fiber of the sphere and on $y$.

\subsubsection*{The solutions}

We are interested in solutions of the action \eqref{6dact} that preserve $SL(2,\mathbb{R})_L \times SU(2)_{R\, or \, L} \times U(1)_{L\, or \, R} \times U(1)_R$ isometry. Using the left/right-invariant forms introduced above, the solution for the metric and the $H$-field takes the general form

\be
ds^2 = \frac{\ell^2}{h} \bigl( - w_+ w_- + \g w_3^2 + \s_1^2 + \s_2^2 + \g \s_3^2 + 2 \g \e_g w_3 \s_3 \bigr)\non
\ee

\be
H = \ell^2 \bigl[ \bigl(\varepsilon \s_1 \wedge \s_2 \wedge \s_3  { -} \half w_+ \wedge w_- \wedge w_3\bigr) +\frac{ \g \,
 \e_B}{h} \bigl(\s_1 \wedge \s_2 \wedge w_3  + \half w_+ \wedge w_- \wedge \s_3 \bigr)\bigr] \label{bck}
\ee
If $\varepsilon = -1$, that is for the $\slr_L \times SU(2)_R$ - invariant Ansatz, we obtain a two-parameter family of solutions \cite{josh}, parameterized by $\e_g$ and $\e_B$. The remaining constants are given by

\be
\g = \frac{1-\e_g^2}{1+\e_B^2 - \e_g^2  {+}\e_B \,\e_g \sqrt{1+\e_B^2 - \e_g^2}}\;, \;\;\;\;\; h = \g \sqrt{1+\e_B^2 - \e_g^2}
\ee
We consider backgrounds where both $\psi$ and $y$ are compact; consequently, absence of closed timelike curves requires that $\g >0$ and $|\e_g| <1$. Noting that one can always choose $\e_g >0$ by an appropriate redefinition of the coordinates, the allowed ranges of the parameters are\footnote{If $y$ is non-compact, then solutions with $\e_g >1$ are also allowed, provided that

\be
\e_B > \sqrt{\e_g^2 -1} \;, \;\;\;\;\; \e_B >0
\ee}

\be
0< \e_g < 1\;, \;\;\;\;\;\; \e_B \in (-\infty, \infty) \label{ranges}
\ee
If $\varepsilon =1$, the equations of motion yield $\e_g = \e_B =0$, and thus the only solutions with $SL(2,\mathbb{R})_L \times SU(2)_L \times U(1)_R^2 $ invariance are locally $AdS_3 \times S^3$. This result does not contradict the fact that there exist dipole ($\e_B \neq 0$) solutions  of type IIB/$M_4$ with $SL(2,\mathbb{R})_L \times SU(2)_L $ symmetry, as the latter require at least two self-dual fluxes to be turned on.

Requiring that the solution be smooth at the poles of the two-sphere spanned by $\th, \phi$ implies that $\psi \sim \psi + 4 \pi$. On the other hand, the quotient of the common D1-D5 direction $y \sim  y + 4 \pi^2 T$ is a free parameter, and corresponds to turning on a right-moving  temperature in the field theory dual \cite{excl}. In this paper we would like to consider more general quotients, which act on $y$ and $\psi$ simultaneously. Denoting these two coordinates by $\hat y^\a$, the coordinate identifications are encoded in the $2 \times 2$ matrix $M$, where

\be
\hat y^\a = M^\a{}_\b\, y^b \;, \;\;\;\;\;\;\; y^\b \sim y^\b + 2 \pi n^\b\;, \;\;\; n^\a \in \mathbb{Z} \label{defm}
\ee
Consequently, $SL(2,\mathbb{R})_L \times SU(2)_R \times U(1)_R \times U(1)_L$-invariant solutions of the action \eqref{6dact} are specified by two nontrivial parameters, $\e_g$ and $\e_B$ satisfying \eqref{ranges}, in addition to the overall scale $\ell$. The global structure of the solution is captured by the matrix $M^\a{}_\b$, which encodes the identifications of the compact coordinates. Note that there exist geometrical restrictions on the possible values of the entries of $M^\a{}_\b$, imposed by the smoothness of the solution at the locations where the fibers degenerate.

\bigskip

As far as supersymmetry is concerned,  it can be checked explicitly that for $\varepsilon =-1$ and $\e_B \neq 0 $ or $\e_g \neq 0$ the dilatino variation equation has no solutions, and thus none of the $\slr_L \times SU(2)_R$ solutions are supersymmetric. For $\varepsilon =1$, only the eight $SL(2,\mathbb{R})_L \times SU(2)_L$ invariant Killing spinors of $AdS_3 \times S^3$ respect the identifications \eqref{defm}, and thus the supersymmetry is broken to half. This analysis agrees with the results of \cite{Gutowski:2003rg}.

\subsection{STU transformations \label{stu}}

All the backgrounds we consider are $T^2$ fibrations over $AdS_2 \times S^2$, where the fibers are $y^\a$. We can reduce along the isometry directions $y^\a$ to four dimensions, to obtain solutions of a $4d$ theory with $SL(2,\mathbb{R})^3$ symmetry known as the STU model\footnote{In the full string theory, the $\slr^3$ symmetry of the supergravity action is broken to $SL(2,\mathbb{Z})^3$. We will mostly discuss the continuous version.} \cite{Duff:1995sm}. An $SO(2,2) \cong SL(2,\mathbb{R})\times SL(2,\mathbb{R})$ part of this symmetry group is nothing but the T-duality group of the compactification two-torus. The remaining $SL(2,\mathbb{R})$ factor is generated by the $Z_2$ transformation \eqref{sduality} combined with one of the $SL(2,\mathbb{R})$ transformations of the compactification torus.

 The six-dimensional fields can be written in Kaluza-Klein form as

\be
ds_6^2 = g_{\mu\nu} \, dx^\mu dx^\nu + G_{\a\b}\, (dy^\a + \A^{\a})(dy^\b + \A^{ \b})
\ee

\be
B  =  (\mathcal{C}_{\mu\nu} - \A^{\a}_\mu \,\B_{\nu \a} + \A^\a_\mu\, B_{\a\b}\, \A^\b_\nu)\, dx^\mu \wedge dx^\nu + 2 (\B_{\mu\a} - B_{\a\b} \A^\b) \, dx^\mu \wedge dx^\a \non
\ee
\be
 \hspace{3 cm} + B_{\a\b} \, dx^\a \wedge dx^\b \label{kkred}
\ee
which yields, from the four-dimensional perspective, Einstein gravity coupled to four gauge fields $A^i = \{ \A^\a, \B_\a\}$ and six scalars descending from $G_{\a\b}, B_{\a\b}, \Phi$ and the four-dimensional Hodge dual of $\mathcal{C}_{\mu\nu}$. The four scalars which descend from the internal metric and B-field naturally parameterize the K\"{a}hler and complex structure of the compactification torus as

\be
G_{\a\b} + B_{\a\b} =  \frac{\rho_2}{\tau_2} \left(\begin{array}{cc} |\tau|^2 & \tau_1 \\ \tau_1 & 1 \end{array}\right) +
\rho_1  \left(\begin{array}{cc} 0 & 1 \\ -1 & 0 \end{array}\right)
\ee
Here $\rho = \rho_1 + i \rho_2$ represents the K\"{a}hler structure parameter of the $T^2$, whereas $\tau = \tau_1 + i \tau_2$ represents the complex structure. The $SO(2,2) \cong SL(2,\mathbb{R}) \times SL(2,\mathbb{R})$ continuous version of the T-duality group thus splits into complex structure transformations

\be
\tau \r \frac{a \tau +b }{c \tau + d} \;, \;\;\;\;\; a d - b c = 1\;, \;\;\;\; \rho \;\; \mbox{fixed}
\ee
and  K\"{a}hler structure transformations

\be
\rho \r \frac{a \rho +b }{c \rho + d} \;, \;\;\;\;\; a d - b c = 1\;, \;\;\;\; \tau \;\; \mbox{fixed}
\ee
The complex structure transformations are simply reparametrizations of the compactification torus which leave its volume unchanged. In the language of the STU model, they correspond to the $\mathcal{U}$ transformations. Given that T-duality can be interpreted as a very simple example of mirror symmetry - which is known to exchange the K\"{a}hler and complex structure of the compactification manifold - it follows that K\"{a}hler structure transformations can be understood as a T-duality, followed by a complex structure transformation and a T-duality back. 
When the parameters of the $\slr$ transformation are $a=d=1$ and $b =0$, which is the situation that will eventually interest us the most, the K\"{a}hler structure transformation corresponds to a  ``TsT'' transformation, or more precisely

\bi
\item a T-duality along $y$
\item a shift $\psi \r \psi + 2 c \, y $
\item a T-duality back along $y$
\ei
Hence, in STU language, K\"{a}hler structure transformations correspond to the $\mathcal{T}$ transformation.

Finally, the $\mathcal{S}$ transformation of the STU model is represented by a fractional linear transformation which acts of the four-dimensional axion-dilaton, with $\rho$ and $\tau$ held fixed. This transformation can be easily obtained by combining a K\"{a}hler transformation with the six-dimensional electromagnetic duality \eqref{sduality}, in the order: electromagnetic duality, K\"{a}hler transformation, electromagnetic duality back. When $M_4 = T^4$, the six-dimensional electromagnetic duality simply corresponds to a ten-dimensional S-duality, four T-dualities on $T^4$ and an $S$-duality back. Thus, we have a quite clear understanding of the string theory interpretation of the STU transformations in this setting, namely type IIB frame with purely NS-NS flux:

\bi
\item $\mathcal{S}$: electromagnetic duality+ K\"{a}hler transformation + electromagnetic duality
\item $\mathcal{T}$: K\"{a}hler transformation (T-duality, coordinate transformation, T-duality back)
\item $\mathcal{U}$: volume-preserving coordinate transformation
\ei

\noindent Note that the three transformations commute, as each of them acts on a different combination of the scalars.
 
\bigskip

In this subsection we would like to show that all the solutions of the action \eqref{6dact} presented in the previous subsection  are related by the above STU transformations. That this should be true was inspired by the fact that in $\N=8$ four-dimensional supergravity there are only two orbits of the U-duality group which relate extremal spherical black hole near-horizons of the form $AdS_2 \times S^2$, one supersymmetric and one non-supersymmetric \cite{Ferrara:1997uz}.

While it is not yet known whether a similar statement holds for $\N=2$ theories \cite{Bellucci:2007ds}, let us assume it is true and understand its implications from a six-dimensional perspective. All four dimensional solutions of the STU model descend from six-dimensional solutions of \eqref{6dact}, which are $T^2$ fibers over a four dimensional base. For the geometries we consider, this base is $AdS_2 \times S^2$. Let us consider the particular fibration giving the full solution $AdS_3 \times S^3$. Before any identifications are made, the $AdS_3 \times S^3$ solution we start from has sixteen Killing spinors, eight left-moving (which depend explicitly only on $\tau,\rho, \th,\phi$) and eight right-moving (which depend only on $y, \psi$) \cite{Duff:1998cr}. Compactifying $AdS_3$ down to $AdS_2$ along $y$ breaks all the right-moving supersymmetries, and this can be seen in two ways: the first is by observing that the original right-moving spinors do not respect the newly-imposed periodicity of the $y$ coordinate; the second is by noticing that in order to make $y$ compact in the near-horizon region one needs to add momentum along this direction, which breaks half the supersymmetries.

We also need to reduce the $S^3$ down to $S^2$. This can be done by making either the $SU(2)_R$ or the $SU(2)_L$ isometry factors manifest, as we have explained in the previous section. If we compactify along $\psi$, the resulting $AdS_2 \times S^2$ background has $SL(2,\mathbb{R})_L \times SU(2)_L$ isometry and inherits all the left-moving Killing spinors from six dimensions. If instead we choose to compactify along $\phi$, the isometry of the resulting  $AdS_2 \times S^2$ is $SL(2,\mathbb{R})_L \times SU(2)_R$, and even though the background still has the eight left-moving supersymmetries, the supersymmetry transformations  will have to involve non-trivial Kaluza-Klein modes on the $T^2$ (since the six-dimensional Killing spinors depend explicitly on $\phi$), and will be completely non-obvious from a four dimensional perspective.

Thus,  the same $AdS_3 \times S^3$ total space gives rise to two inequivalent (i.e. not related by an STU transformation) $AdS_2 \times S^2$ four-dimensional backgrounds, one with
$SL(2,\mathbb{R})_L \times SU(2)_L$ isometry, and the other with  $SL(2,\mathbb{R})_L \times SU(2)_R$ symmetry. The STU transformations associated with the explicitly $SL(2,\mathbb{R})_L \times SU(2)_L$ invariant reduction  will always preserve the eight Killing spinors and generate the supersymmetric orbit(s), while the STU transformations which respect $SL(2,\mathbb{R})_L \times SU(2)_R$ will generate the non-supersymmetric one(s). In the language of the previous section,  the supersymmetric orbit(s) consist of the six dimensional backgrounds with $\varepsilon =1$, all of which are locally $AdS_3 \times S^3$, whereas the non-supersymmetric ones contain the $\varepsilon =-1$ backgrounds. We have checked explicitly that in the STU model there is only one orbit of each kind, and thus all $\varepsilon =-1$ solutions with different values of $\e_B, \e_g$  are related by STU transformations\footnote{We thank A. Strominger for insightful discussions of this point.}.

\subsubsection*{K\"{a}hler transformations of the compactification $T^2$}

We start from $AdS_3 \times S^3$, characterized by the string-frame length $\ell$, $\e_B=\e_g =0$ and identification matrix $M_0$. We perform a K\"{a}hler transformation with parameter

\be
\Lambda = \left(\begin{array}{cc} a & b \\ c & d \end{array}\right)\;, \;\;\;\;\; a d - bc =1
\ee
Using the formulae in  appendix \ref{so22}, we find that after the transformation

\be
\e_B =  {-} \e_g =  x \;\;\;\;\;\; \mbox{if} \; \varepsilon = - 1	
\ee {
where we introduced the shorthand

\be
x \equiv  \frac{2 c d \V}{c^2 \V^2 + d^2}
\ee
and let $\V = \ell^2 \det M_0$. On the other hand,

\be
\e_g = \e_B =0 \;\;\;\;\;\; \mbox{if} \; \varepsilon =  1
\ee
This result agrees with the fact that for $\varepsilon = 1$ all solutions of the equations of motion are locally $AdS_3 \times S^3$. The flux through the three-sphere, $\ell^{2}$, is unchanged. The coordinate identifications following this transformation are given by

\be
M_1 = \frac{d {+} c \V  \s_1}{d^2 - c^2 \V^2 }\,  M_0 \;\;\;\;\;\; \mbox{if} \; \varepsilon = -1	
\ee
and

\be
M_1 = \frac{d  {-} c \V \, \e}{d^2 + c^2 \V^2 }\, M_0 \;\;\;\;\;\; \mbox{if} \; \varepsilon =  1	
\ee
Thus, starting from a diagonal $M_0$ we generically end up with one which has off-diagonal terms. The new six dimensional dilaton is given by

\be
e^{2\Phi_1} = \frac{1}{d^2 + c^2 \V^2} e^{2 \Phi_0}
\ee

\subsubsection*{The electromagnetic transformation}

The action of the six-dimensional electromagnetic duality on the various fields is given in \eqref{sduality}.
From now on we will only work with the $\varepsilon = - 1$ backgrounds, as no new $\varepsilon =1$ backgrounds can be generated through the transformations that follow\footnote{The reason is that all  geometries with $\varepsilon =1$ are locally $AdS_3 \times S^3$, and that even at the level of identifications, the effect of a combination of T and S transformations is entirely reproducible by just a T transformation with an appropriately chosen parameter.}. Using the formulae in appendix \ref{hodge}, we find that the new background has

\be
\e_g' = \e_g = {-} x \;, \;\;\;\;\;\; \e_B' =0
\ee
and radius 

\be
L^2 = e^{-2 \Phi_1} \ell^2  = (d^2 -c^2 \V^2) e^{-2 \Phi_0} \ell^2
\ee
Since the electromagnetic duality acts at the level of the field strength, we can simply assume that after the duality there is no constant $B$-field in the internal directions. Thus, the resulting K\"{a}hler parameter is 

\be
\rho_1'=L^2 \frac{\g' \e_B'}{h'} + \Delta B' =0 \;, \;\;\;\;\; \rho_2'= \frac{L^2 \g' }{h'} \sqrt{1-\e_g^2} \det M_1 = L^2 \det M_1 \equiv \tilde{\V}
\ee
where 

\be
\tilde{\V} = \V \, e^{-2\Phi_0} \label{relvol}
\ee

\subsubsection*{K\"{a}hler transformation after the electromagnetic duality}

Now, we take the background we have just described and perform one more K\"{a}hler transformation on it, given by

\be
\tilde{\Lambda} = \left(\begin{array}{cc} \;\tilde{a}\; &\; \tilde{b}\; \\ \tilde{c} & \tilde{d} \end{array}\right)\;, \;\;\;\;\; \tilde{a} \tilde{d} - \tilde{b} \tilde{c} =1
\ee
The parameters of the resulting background are

\be
\tilde{\e}_g = - \frac{x + \tilde{x} }{1 + x \,\tilde{x}} \;, \;\;\;\;\;\;\; \tilde{\e}_B = \,\frac{\tilde{x}\sqrt{1-x^2}}{1 + x \, \tilde{x} }
\ee
where we have defined

\be
\tilde{x} \equiv \frac{2 \tilde{c} \tilde{d} \tilde{\V}}{  \tilde{d}^2+\tilde{c}^2 \tilde{\V}^2}
\ee
The other quantities of interest are 

\be
\tilde{\g} = 1 + x \, \tilde{x} \;, \;\;\;\;\; \tilde{h} = h = \sqrt{1-x^2}\;, \;\;\;\;\; \tilde{L} = L
\ee
All allowed values of $\tilde{\e}_g \in (-1,1)$ and $\tilde{\e}_B \in (- \infty, \infty)$ are obtainable by this set of transformations, provided $\Lambda, \tilde \Lambda \in \slr$. In particular, the non-supersymmetric dipole backgrounds\footnote{We define the dipole backgrounds to be those solutions to \eqref{6dact} which have $\e_g=0$ and $\e_B \neq 0$. They are related by an $SO(5,21)$ U-duality to the usual three-dimensional dipole backgrounds. } are obtained by setting $x + \tilde{x} =  0$.

The value of the dilaton after all these transformations is 

\be
e^{2\Phi_2} = \frac{1}{\tilde{c}^2 \tilde{\V}^2+\tilde{d}^2} \, e^{2 (- \Phi_1)} = \frac{c^2 \V^2 + d^2}{\tilde{c}^2 \tilde{\V}^2+\tilde{d}^2} e^{-2 \Phi_0}
\ee
while the identifications and the constant internal B-field shift are given by

\be
M_2 =  \frac{1}{\tilde{d}^2 - \tilde{c}^2 \tilde{\V}^2}\left(\begin{array}{cc}\;\; \tilde{d} & \;\; \;\tilde{c}\tilde{\V}\\ \tilde{c}\tilde{\V} &\;\; \tilde{d} \end{array}\right)\,M_1
\;, \;\;\;\;\;\;  \Delta B_2 = L^2 \left(\frac{\tilde{b} \tilde{d}}{\tilde{\V}}-\tilde{a} \tilde{c}\, \tilde{\V}\right)
\ee

\subsubsection*{The final electromagnetic transformation}

We can perform one last electromagnetic duality on the background, such that we can encompass the last three transformations as an $\mathcal{S}$ transformation in the STU sense. Its effect is simply to interchange $x$ and $\tilde{x}$ in the final expressions, so we now have a background with 

\be
\tilde{\e}_g' = - \frac{ x+ \tilde{x} }{1+ x \, \tilde{x} } \;, \;\;\;\;\;\;\; \tilde{\e}_B' = \frac{x \sqrt{1-\tilde{x}^2}}{1 + x \, \tilde{x} }
\ee
and $\tilde{\g}' = \tilde{\g}$. The final dilaton is

\be
e^{2 \Phi_2'} = \frac{\tilde{c}^2 \tilde{\V}^2+\tilde{d}^2}{c^2 \V^2 + d^2} \, e^{2 \Phi_0}
\ee
whereas the final string frame radius of the geometry is 
\be
\tilde{L}^{'2} = e^{-2 \Phi_2} L^2 = (\tilde{d}^2-\tilde{c}^2 \tilde{\V}^2) \ell^2
\ee
By applying a $\mathcal{U}$ transformation one can also bring the identifications matrix $M_2$ into a desirable form. We will make use of this last transformation in the next section, for the specific task of matching to the near-horizon geometry of the non-supersymmetric extremal five-dimensional Kerr-Newman black hole.

\subsection{Matching to NHEK}

In the previous subsection we have shown that $\slr_L \times SU(2)_R$ invariant backgrounds with arbitrary $\e_g, \e_B$ parameters can be generated via a sequence of $\mathcal{S}$ and $\mathcal{T}$ transformations. It has been long known \cite{Dias:2007nj} that the near horizon geometry of the six-dimensional uplift of the non-supersymmetric extremal rotating D1-D5-p black hole (NHEK) is a geometry of the type \eqref{bck}. In this subsection we review the relation between the charges of the original D1-D5-p black hole and the $\e_{g,B}, \Phi, \ell, M^\a{}_\b$ parameters characterizing its near-horizon geometry and find the parameters of the STU transformations that map $AdS_3 \times S^3$ into six-dimensional NHEK.

We consider the extremal non-supersymmetric D1-D5-p black hole with charges $Q_1$, $Q_5$, $Q_p$ and  left-moving angular momentum $J_L$ \cite{Breckenridge:1996sn,Cvetic:1996xz}. Its charges can be parameterized as

\be
Q_1 = 2 a^2 \sinh 2 \d_1\;, \;\;\;\;\; Q_5 = 2 a^2 \sinh 2 \d_5 \;, \;\;\;\;\; Q_p = 2 a^2 \sinh 2 \d_p
\ee

\be
J_L = 4 a^3 (\cosh \d_1 \cosh \d_5 \cosh \d_p + \sinh \d_1 \sinh \d_5 \sinh \d_p)
\ee
and its entropy is given by

\be
S= 2 \pi \sqrt{J_L^2 - Q_1 Q_5 Q_p} \label{entkerr}
\ee
The uplift to six dimensions of the near-horizon geometry of this black hole can be written in terms of $\slr_L \times SU(2)_R$
invariant forms as in \eqref{bck}, where the various parameters are given by

\be
\g \,\e_g = \frac{1}{\cosh 2 \d_1} + \frac{1}{\cosh{2 \d_5}} \;, \;\;\;\;\;\;\;\; \g \,\e_B=  {-} \frac{\tanh 2 \d_5}{\cosh 2 \d_1} \label{ekerr}
\ee
The identifications are

\be
y \sim y + 4 \pi^2 T_Q m \;, \;\;\;\;\; \psi \sim \psi + 4 \pi n - \frac{4\pi J_L m}{Q_1 Q_5}
\ee
with

\be
T_Q = \frac{J_L T_R}{Q_1 Q_5}  \;, \;\;\;\;\; \pi T_R = \sqrt{1 - \frac{Q_1 Q_5 Q_p}{J_L^2}}
\ee
Thus, the above identifications can be encompassed in the matrix

\be
M_{NHEK} = \left(\begin{array}{cc} 2 \pi T_Q & \;\;0\;\; \\ - \frac{2J_L}{Q_1 Q_5} & 2\end{array}\right)
\ee
In order to compare this geometry to the one in the previous section we need to perform a type IIB S-duality, which turns the three-form RR flux which supports this geometry into NS-NS flux. In the new string frame, the flux of $H^{(3)}$ through the three-sphere and six-dimensional dilaton are

\be
L^2_{NHEK} = \frac{a^2}{2} \sinh 2 \d_5\;, \;\;\;\;\;  e^{2\Phi_{NHEK}} = \frac{\cosh 2 \d_5}{\cosh 2 \d_1}
\ee

\bigskip

\noindent The above represent a complete set of data characterizing the NHEK geometry. Our task now is to match this data to what we obtain by performing an STU transformation on $AdS_3\times S^3$. The three STU matrices are parameterized as

\be
\mathcal{S} = \left(\begin{array}{cc} \tilde a & \;\;\tilde b\; \\ \tilde c & \; \tilde d \end{array} \right) \;, \;\;\;\;\;\;\mathcal{T} = \left(\begin{array}{cc} a & \;\; b\; \\ c & \;d \end{array} \right) \;, \;\;\;\;\;\;\mathcal{U} = \left(\begin{array}{cc} a' & \;\; b'\; \\ c' & \;d' \end{array} \right) \label{stumat}
\ee
where the unimodularity condition holds for each. The input $AdS_3 \times S^3$, in the NS-NS frame,  is taken to be the near horizon geometry of a stack of $q_5$ NS5 branes and $q_1$ F1-strings, carrying momentum $q_0$. Thus, the input data is

\be
\ell^2 = \frac{q_5}{4} \;, \;\;\;\;\; e^{2\Phi_0} = \frac{q_5}{q_1} \;, \;\;\;\;\; M_0 = \left(\begin{array}{cc} 2 \pi T_0 & 0 \\ 0 & 2 \end{array} \right)
\ee
where the right-moving temperature $T_0$ is given by

\be
T_0 = \frac{1}{\pi} \sqrt{\frac{q_0 }{q_1 q_5}}
\ee
The entropy carried by the original black string is

\be
S = 2 \pi \sqrt{q_1 q_5 q_0} \equiv 2 \pi \hat S_0
\ee
The details of the matching are presented in appendix \ref{match}. It is quite clear that, given any final desired $Q_1, Q_5, Q_p$ and $J_L$, one can always find matrices $\mathcal{S},\mathcal{T}, \mathcal{U} \in SL(2,\mathbb{R})$ which map $AdS_3 \times S^3$ to NHEK for some choice of  input data $q_1, q_5, q_0$. In fact, as we show in appendix \ref{match}, the matching conditions leave two parameters unspecified, which we can choose to be $d$ and $\tilde{d}$ in \eqref{stumat}.

The question is, then, whether we can restrict the form of the $\mathcal{S}, \mathcal{T}, \mathcal{U}$ matrices in physically interesting ways. An obvious question is whether the three transformations can be full string dualities, rather than just supergravity ones, that is if we can find  $\mathcal{S},\mathcal{T},\mathcal{U} \in SL(2, \mathbb{Z})$. If such transformations existed, then it would mean that the theory dual to NHEK is U-dual to the D1-D5 CFT. Nevertheless, we show in appendix \ref{match} that the $\mathcal{T}$ and $\mathcal{S}$ transformation matrices cannot both be integer-valued.

Another interesting way to restrict the $SL(2,\mathbb{R})^3$ transformations is by requiring that the matrices have $a=d=1$ and $b=0$. As we will explain in the next section, the $\mathcal{S}$ and $\mathcal{T}$ transformations have a simple interpretation in terms of string/M-theory dualities - and a possibly tractable microscopic description - if the matrices take this particular form. We can apply these restrictions to the $\mathcal{S}$ and $\mathcal{T}$ transformations but, as it can be seen from \eqref{finu}, the $\mathcal{U}$ transformation has to be left general (because $b'>0$). A simple example is worked out below.

\subsubsection*{An example}

 To give the reader a rough idea about how the match works, we present herein the simplest example:  the STU transformations
that map $AdS_3 \times S^3$ to the $6d$ NHEK geometry with equal charges $Q_1=Q_5=Q_p=Q$ and angular momentum $J_L$, assuming the first two transformations are ``TsT''s. As we show in the appendix, the input geometry must have

\be
q_1=q_5 =q \;, \;\;\;\;\;\; q_0 = \frac{J_L^2 - Q^3}{q^2}
\ee
where $q$ is the real, positive, solution to the cubic equation\footnote{Interestingly, the mass of the extremal non-supersymmetric D1-D5-p black hole satisfies a similar equation. The parameter $q$ above is related to $M$ via
\be
q = \frac{M}{6} + \frac{Q}{2}
\ee}

\be
(q-Q)(4 q - Q)^2 = J_L^2 - Q^3
\ee
In addition, the parameter $q$ needs to be integer, given that both $Q$ and $J_L$ are. The central charge of the initial CFT is $c=6 q^2$. The only non-trivial parameters of the  $\mathcal{T}$ and $\mathcal{S}$ transformations are

\be
c = \tilde c =  - \frac{\sqrt{q} }{4 q -Q }
\ee
and of the $\mathcal{U}$ one

\be
a' = \frac{q}{Q^2} \left(2 q -Q - \frac{2 J_L \sqrt{q}}{4 q -Q} \right) \;, \;\;\;\;\;\; b' = \frac{2 q^{3\over 2}}{4 q -Q}
\ee

\be
c'= \frac{2 (4 q -Q) (q-Q)}{Q^2 \sqrt{q}} - \frac{2q -Q}{q Q^2} J_L \;, \;\;\;\;\; d' = \frac{2 q - Q}{q}
\ee

\section{Spectral flows and Melvin  twists \label{spflmel}}

In this section we begin by discussing spectral flows and generalized spectral flows, and their action on our geometries. We then show that the STU duality transformations or the STsTS transformations that take one from $AdS_3 \times S^3$ to $6d$ NHEK are in fact a combination of two generalized spectral flows. We also observe that the action of generalized spectral flows on six-dimensional geometries with D1 and D5 charges and two commuting isometries has a very simple geometrical interpretation, and use this to write down a straightforward formula for the action of generalized spectral flows on such geometries. We also show that these generalized spectral flows (or STsTS transformations) can be seen as a combination of T-dualities and Melvin twists.

\subsection{Spectral flows and their generalizations}

The bosonic generators of the small $\N=(4,4)$ superconformal algebra in two dimensions are the Virasoro generators $L_n$ and the $SU(2)$ Ka\v{c}-Moody current algebra generators $J^a_n$, where $a$ is an $SU(2)$ index, together with their right-moving counterparts. This algebra has an $SU(2)$ (inner) automorphism \cite{seibschwimm} known as spectral flow, which shuffles the left-moving  generators as

\be
L_n \r L_n + \eta J^3_n + \frac{c}{6}\, \eta^2 \d_{n,0} \;, \;\;\;\;\; J_n^3 \r J_n^3 + \frac{c}{3} \,\eta\, \d_{n,0} \label{specfl}
\ee
and similarly on the right. These spectral flow symmetries have a simple geometrical description \cite{Balasubramanian:2000rt,krauslarsen, kraushansen} in the context of the $AdS_3/CFT_2$ correspondence, in which the gravitational background dual to the $(4,4)$ SCFT is $AdS_3 \times S^3$. Namely, they are represented by large diffeomorphisms which mix the the sphere coordinates with the boundary coordinates of $AdS_3$. These coordinate transformations take the form

\be
\phi \r \phi + \l\, z\;, \;\;\;\;\;\psi \r \psi + \bar \l \, \bar z \label {spflsptz}
\ee
where the parameter $\l$ of the diffeomorphism in spacetime is related to the left/ right spectral flow parameter by $\l = 2 \eta$, $\bar \l = 2 \bar \eta$ and the correspondence between the Euclideanized boundary coordinates ($z, \bar z$) and the Lorentzian light-like ones ($y,t$) is $z \r t$ and $\bar z \r y$. The angular momenta $\p_\phi, \p_\psi$ represent the zero-modes of the $J^3$ component of the $SU(2)_L$ and  $SU(2)_R$ R-symmetry currents, respectively. The relationship between the gauge transformations \eqref{spflsptz} and the CFT automorphisms \eqref{specfl} is nicely explained in \cite{krauslarsen} in terms of the holographic dictionary.

It is interesting to remark that the large diffeomorphisms we consider in this paper (for $\varepsilon =-1$) are
of the form

\be
\phi \r \phi + \l \, \bar z \label{nonspfl}
\ee
rather than \eqref{spflsptz}. Using the holographic dictionary for $SU(2)$ Chern-Simons gauge fields in $AdS_3$, one finds that the above transformation corresponds to deforming the original CFT action $S_\star$ by the left-moving current

\be
S = S_\star + \l \int d^2 z \, J_L^3(z)
\ee
This deformation can be absorbed into a redefinition of the operators of the CFT, and bears no effect on the R-charges or conformal dimensions. Therefore, it does not correspond to a CFT spectral flow. Nevertheless, we will still use the terminology of spectral flows, even for diffeomorphisms of the type \eqref{nonspfl}.

\bigskip

The spectral flow diffeomorphisms have been also widely used as solution-generating techniques of five-dimensional asymptotically locally flat geometries. The idea is roughly as follows. In string theory, the $AdS_3 \times S^3$ geometry arises in the near-horizon limit of a stack of parallel D1 and D5 branes, where the D5's are additionally wrapping an internal four-dimensional manifold $M_4$. In general, there is also (right-moving) momentum along the common direction of the D1 and D5 branes - call it $y$ - and the entire configuration is BPS. If we imagine the coordinate $y$ to be compactified, the asymptotics of this solution are $\mathbb{R}^{1,4} \times S^1$. From a five-dimensional perspective - obtained by Kaluza-Klein reduction along $y$ - the D1-D5-P black string becomes a three-charge asymptotically flat black hole.

The full solution has an $S^3$ factor all the way from the near-horizon $AdS_3 \times S^3$ to the asymptotic spatial $\mathbb{R}^4$ (the latter has an $S^3$ factor when written in spherical coordinates), and hence has $SO(4)$ symmetry. Let $\psi$ for now denote either the $\psi$ or the $\phi$ angle of the $S^3$. Since both $\psi$ and $y$ are isometry directions throughout the solution, one may try to investigate the effect of the spectral flow transformation

\be
\psi \r \psi + \l \, y \label {spflspt}
\ee
on the \emph{full} geometry. Asymptotically, the metric is $\mathbb{R}^{1,4} \times S^1_y \times M_4$

\bea
ds^2_{r \r \infty} & = & - dt^2  + dy^2   + dr^2 + r^2 d\Om_3^2 + ds_M^2  \non \\
&=& - dt^2 + dy^2 + \rho\, (d\psi + \cos\th\, d\phi)^2 + \rho\, (d\th^2 + \sin^2\th \, d\phi^2) + \frac{d\rho^2}{ \rho} + ds_{M}^2
\label{asfl}\eea
where we have defined a new radial coordinate $\rho = \frac{r^2}{4}$. Applying the transformation \eqref{spflspt} and then reducing to five dimensions along $y$ one obtains a linearly diverging dilaton and a geometry with a singular behavior as $\rho \r \infty$. Nevertheless, if one first embeds the D1-D5-p system in Taub-NUT space and aligns the Taub-NUT circle with $\psi$, this problem is no longer encountered\cite{Bena:2008wt}; the asymptotics of the solution are instead $\mathbb{R}^{1,3} \times S^1_\psi \times S^1_y$

\be
ds^2_{TN} = - dt^2 + dy^2 + V^{-1} (d\psi + \cos\th d\phi)^2  + V\left( d \rho^2 + \rho^2  (d\th^2 + \sin^2\th d\phi^2)\right)  + ds_{M}^2
\ee

\be
V(\rho) = 1 + \frac{1}{\rho} \;, \;\;\;\; \lim_{\rho \r \infty } V(\rho) = 1
\ee
and the effect of the ``global'' spectral flow asymptotically takes the form

\be
ds^2_{sp.fl.} = (1+\l^2) \left(dy + \frac{\l}{1+ \l^2}(d\psi + \cos\th d\phi) \right)^2  + \frac{1}{1+ \l^2} (d\psi + \cos\th d\phi)^2  + ds^2_{\mathbb{R}^{1,3}} + ds_{M}^2
\ee
Thus, from the perspective of five-dimensional supergravity, the geometry is still asymptotically $\mathbb{R}^{1,3} \times S^1_\psi$, and only the asymptotic values of the dilaton and the Kaluza-Klein gauge field change. Note that while this is a rather non-trivial modification from a five-dimensional perspective, in the six-dimensional picture it corresponds to a simple diffeomorphism. Note also that the embedding in Taub-NUT space does not affect the near-horizon geometry of the D1-D5 system, since close to its center, Taub-NUT is indistinguishable from $\mathbb{R}^4$.

\bigskip

Let us now discuss the generalization of ``global'' spectral flows introduced in \cite{Bena:2008wt}. The supergravity fields sourced by the D1-D5 configuration are solution to a simpler $6d$ consistent truncation of the type IIB action containing only the metric, the RR two-form potential $C^{(2)}$ with field strength $F=d C^{(2)}$ and a scalar\footnote{The scalar $e^{2\tilde{\Phi}}$ is the string frame volume of $M_4$.} $\tilde{\Phi}$ 

\be
S_{RR} = \int d^6 x \sqrt{g} \left( R - (\p \tilde \Phi)^2 - \frac{1}{12} e^{2 \tilde \Phi} F^2  \right)
\ee
Upon dimensional reduction to five dimensions, this action yields $\N=2$ five-dimensional supergravity coupled to two vector multiplets. The bosonic content of this  theory consists of the metric, three  gauge fields $A^{(i)}$ and two scalars. One of the gauge fields is the Kaluza-Klein gauge field one obtains by reducing the metric, another  one comes from the reduction of the $C^{(2)}$ field, and the third is the five-dimensional Hodge dual of the RR two-form.

\be
A^{(1)}_\mu = \frac{g_{\mu y}}{g_{yy}} \;, \;\;\;\;\; A^{(2)}_\mu = C^{(2)}_{\mu y} \;, \;\;\;\;\;  d A^{(3)} = \star_5 d C^{(2)}
\ee
As already discussed, the spectral flow \eqref{spflspt} has a non-trivial action on the five-dimensional fields, whereas it is a simple coordinate transformation in the six-dimensional space-time

\be
ds_6^2 = ds_5^2 + g_{yy} \,(dy + A^{(1)}_\mu dx^\mu)^2\,.
\ee
The observation of \cite{Bena:2008wt} was that the above five-dimensional model has a discrete symmetry that interchanges the three gauge fields. Consequently, one should be able to have a spectral flow associated to a diffeomorphism of the U-dual six-dimensional metric

\be
ds^2_{6'} = ds_5^2 + g_{yy}' (dy' + A^{(2)}_\mu dx^{\mu})^2
\ee
in which $A^{(2)}$ plays the role of Kaluza-Klein gauge field, and similarly for $A^{(3)}$. One thus obtains two ``generalized spectral flows'', which are very useful solution-generating techniques in five dimensions. In the following, we will give a simple interpretation of these transformations in terms of string dualities.

The simplest way to note that the five-dimensional theory has a permutation symmetry is to obtain it not from a truncation of type IIB on $M_4 \times S^1$ for $M_4=T^4$, but rather as one of M-theory on $T^2 \times T^2 \times T^2$. The three gauge fields descend from the  M-theory three-form potential with two components along a given $T^2$, and the permutation symmetry between them is manifest.

Nevertheless, even in the D1-D5 type IIB frame this symmetry is not hard to understand. If we perform an S-duality on the system, we obtain a solution of type IIB supergravity on $M_4$ with only NS-NS flux

\be
S_{NS-NS} = \int d^6 x \sqrt{g} \left( R -  (\p \Phi)^2 - \frac{1}{12} e^{-2 \Phi} H^2  \right)
\ee
where $H=dB$ is the Neveu-Schwarz three-form field and $\Phi$ is the six-dimensional dilaton. The reduction to five dimensions works as before, with $C^{(2)}$ replaced by $B^{(2)}$. Now suppose we want to interchange the $5d$ gauge fields $A^{(1)}$ and $A^{(2)}$. Given the equivalence of type IIA string theory compactified on a circle with type IIB on $S^{1}$, it is trivial to see that the necessary transformation is a T-duality in the $y$ direction. Consequently, the first generalized spectral flow can be obtained from the following sequence of transformations:

\be
\mbox{STsTS} = \mbox{S-duality}, \;\; \mbox{T-duality along}\; y , \; \;\mbox{shift:} \; \psi \r \psi + \l\, y , \;\; \mbox{T-duality on}\; y, \;\;\mbox{S-duality} \non
\ee
The second generalized spectral flow, obtained by interchanging $A^{(1)}$ with $A^{(3)}$, is obtained by first performing four T-dualities on $M_4$, an operation which implements Hodge duality in six dimensions ($F_3 \r \star_6 F_3$) and exchanges $A^{(2)}$ and $A^{(3)}$ from the five-dimensional perspective, followed by the same transformations as before. Consequently, the second generalized spectral flow can be summarized as: $T^4 S \,TsT\, S T^4$. In the next section, we will use this definition in terms of dualities to obtain the action of each spectral flow on a generic six-dimensional metric with two compact isometry directions.

\bigskip

Before ending this section, let us make a few comments concerning supersymmetry. The D1-D5-p solution preserves four supersymmetries, and the Taub-NUT solution sixteen. Four supersymmetries can be preserved in the juxtaposition assuming that the Taub-NUT space is properly aligned with respect to the isometries of the D1-D5-p solution. More concretely, let us assume that the momentum is right-moving, so from a near-horizon perspective the supersymmetries of the D1-D5-p system are associated with $SU(2)_L$ only. If the Taub-NUT circle is aligned with $\psi$, then the entire solution preserves $SU(2)_L$, and supersymmetry is preserved throughout. Nevertheless, if the Taub-NUT  circle is $\phi$, then $SU(2)_L$ is not preserved in the full solution and supersymmetry is broken. At the level of the metric, the two embeddings differ by a trivial interchange of coordinates $\phi \leftrightarrow \psi$, and the only difference, which is responsible for supersymmetry breaking, occurs in a relative sign in the two-form gauge potential. Thus, the non-supersymmetric solution is also called almost-BPS \cite{Goldstein:2008fq}.
The sign difference becomes important, nevertheless, when performing generalized spectral flows. The BPS solutions remain BPS, and the generalized spectral flows just act by interchanging the eight harmonic functions \cite{Gauntlett:2004qy,Bena:2005ni,Bates:2003vx} underlying these solutions  \cite{Bena:2008wt}. However, the almost-BPS solutions are transformed into solutions that do not belong to the almost-BPS class; for example, one spectral flow gives a solution with an Israel-Wilson base \cite{Bena:2009fi}, and three generalized spectral flows give the rather complicated solutions of \cite{DGR}.

\subsection{The spectral-flowed geometries \label{spflgeom}}

In this section we derive the effect of the generalized spectral flows we have discussed, namely $S\,TsT\,S$ and $T^4 S\,TsT\, ST^4$, on an arbitrary type IIB background supported by purely RR three-form flux. The only requirement is that the background in question have two compact, commuting isometries, so that the metric can be written as a $T^2$ fibration over an eight-dimensional base

\be
ds_{10}^2 = ds_8^2 + G_{\a\b}\, (dy^\a + \A^\a)(dy^\b + \A^\b)  \;, \;\;\;\;\; y^\a \sim y^\a + 2 \pi \label{metfib}
\ee
In addition, we assume that only the dilaton $\phi$ and the two-form potential are turned on. The latter can be decomposed as

\be
C^{(2)}_{\a\b} = \zeta \hat\e_{\a\b} \;, \;\;\;\;\; C^{(2)}_{\mu\a} = \B_{\mu\a} - C_{\a\b} \A^\b \non
\ee

\be
C_{\mu\nu}^{(2)} = \mathcal{C}_{\mu\nu} - \A_{[\mu}^\a\, \B_{\nu] \a} + \A^\a_\mu\, C_{\a\b}\,  A^\b_\nu \label{decc}
\ee
The matrix $\hat{\e}_{\a\b} = i \s_2$ represents the two-dimensional $\e$ symbol, whereas the unhatted $\e_{\a\b} = \sqrt{G}\, \hat{\e}_{\a\b}$ is the corresponding tensor density. Let us now study the effect of the first generalized spectral flow on the above generic field configuration.

\subsubsection*{The first generalized spectral flow: $S\,TsT\,S$}

After a type IIB S-duality, we obtain

\be
\phi' = - \phi \;, \;\;\;\;\; ds_{10'}^2 = e^{-\phi} ds_{10}^2 \;, \;\;\;\;\; B_{MN}= C_{MN}
\ee
Now we perform a TsT transformation on the resulting metric. As we have already discussed, this TsT transformation is obtained as a K\"{a}hler transformation on the $T^2$, with parameter\footnote{In terms of the spectral flow parameter in \eqref{spflspt}, $\hat \l_1 = \half \l_1$. The factor of $\half$ takes care of the fact that the periodicity of the Hopf fiber coordinate $\psi$  is $4 \pi$ rather than $2\pi$. }

\be
\Lambda_1 = \left(\begin{array}{cc} \;1\; &\; 0\; \\ \hat \l_1 & 1 \end{array} \right)
\ee
Consequently, after the transformation we have

\be
\A^{''\a} = \A^\a - \hat \l_1 \, \hat \e^{\a\b} \, \B_\b \;, \;\;\;\; \B''_\a = \B_\a
\ee

\be
G_{\a\b}'' = \frac{G'_{\a\b}}{ (1+ \hat \l_1 \,\zeta)^2 + \hat \l_1^2 \det G'}\, \;\;\;\;\; \zeta'' = \frac{\zeta + \hat \l_1 (\zeta^2 + \det G')}{ (1+ \hat \l_1 \,\zeta)^2 + \hat \l_1^2 \det G'}
\ee
The new dilaton is

\be
e^{2 \phi''} = \frac{e^{2 \phi'}}{(1+ \hat \l_1 \,\zeta)^2 + \hat \l_1^2 \det G'}
\ee
The last step is to perform an S-duality back, yielding

\be
ds^{2'''}_{10} = \Sigma_1^\half\, ds_8^2 + \Sigma_1^{-\half} G_{\a\b}\, (dy^\a + \A^\a- \hat \l_1 \, \hat\e^{\a\g} \, \B_\g)(dy^\b + \A^\b- \hat \l_1 \, \hat\e^{\b\g} \, \B_\g) \label{tstmet}
\ee

\be
e^{2 \phi'''} =  e^{2 \phi} \Sigma_1 \;, \;\;\;\;\; \zeta''' =  \frac{\zeta + \hat \l_1 (\zeta^2 + e^{-2 \phi} \det G)}{\Sigma_1}\;, \;\;\;\;\; \B_\a'''=  \B_\a\;, \;\;\;\;\; \mathcal{C}_{\mu\nu}''' =  \mathcal{C}_{\mu\nu} \label{newparamone}
\ee
where we have defined

\be
\Sigma_1 =  (1+ \hat \l_1 \, \zeta)^2 + \hat \l_1^2 e^{-2\phi} \det G
\ee
For future purposes, let us define the following  $n$-forms $H^{(n)}$ on the eight-dimensional base 
\be
H^{(1)}_{\a\b} = d\, C_{\a\b} \;, \;\;\;\;\; H^{(2)}_\a = d \B_\a - C_{\a\b}\, d \A^\b
\ee

\be
 H^{(3)} = d\, \mathcal{C} -\half\, \A^\a \wedge d \B_\a - \half \,\B_\a \wedge d \A^\a \label{defh3}
\ee
The transformation of the internal $C$-field, and consequently of $H^{(1)}$, can be rewritten as

\be
\zeta''' = - \frac{\zeta}{\Sigma_1} + \frac{1}{\hat \l_1} \left(1- \frac{1}{\Sigma_1}  \right)
\ee
The transformation of the  forms $H^{(n)}$ under $STsTS$ is

\be
H^{(3)} \r H^{(3)} \;, \;\;\;\;\; H^{(2)}_\a \r \frac{1+ \hat \l_1 \zeta}{\Sigma_1} H^{(2)}_\a - \frac{\hat \l_1 e^{-2 \phi} \det G}{\Sigma_1}\, \hat \e_{\a\b} d \A^\b \label{trh}
\ee
It is not hard to check that the four-dimensional dilaton (in the NS frame) and  axion (the four-dimensional Hodge dual of $\mathcal{C}_{\mu\nu}$), as well as the complex structure of the compactification torus, are unchanged. 
 Consequently, the first generalized spectral flow is represented by the $\mathcal{T}$ transformation of the STU model.

\bigskip

\subsubsection*{The second generalized spectral flow: $T^4S\,TsT\,ST^4$}

Now let us study the effect of the other generalized spectral flow. We start again from \eqref{metfib}, but now we need to make the additional assumption that the eight-dimensional base geometry is the direct product of a compact $T^4$ and a non-compact piece

\be
ds_8^2 = ds_4^2 + ds_M^2 \;, \;\;\;\;\;\; ds_M^2 = v(x^\mu) \sum_{i=1}^4 dx_i^2   \;, \;\;\;\;\; x^i \sim x^i + 2 \pi
\ee
and that the $C^{(2)}$-field does not have any components in the internal directions of $M_4$, i.e. that $\mu, \nu$ now run only from one to four. Performing four T-dualities on $M_4$ we obtain

\be
ds_8^{'2} = ds_4^2 + v^{-1}  \sum_{i=1}^4 d x_i^2 \;, \;\;\;\;\;\; C^{'(6)} = C^{(2)} \wedge \prod_{i=1}^4  dx^i\;, \;\;\;\;\; e^{-\phi'} = v^2 e^{-\phi}
\ee
The next step is to perform an $S\, TsT\, S$ transformation on this geometry, for which we can use the formulae derived in the previous subsection. Nevertheless, we have to first perform a Hodge duality to turn $C^{'(6)}$ into a two-form $C^{'(2)}$, which can then be decomposed according to \eqref{decc}.  The field strength associated to $C^{'(2)}$ is related to that of $C^{(2)}$ via

\be
F_3' =v^2 \star_6 F_3 \;, \;\;\;\;\; F_3 = d C^{(2)}
\ee
where the six-dimensional space consists of the four non-compact directions and the $T^2$ fiber. From the point of view of four dimensions, Hodge duality is rather trivial, in the sense that it just exchanges the forms $H^{(n)}$ introduced earlier. More precisely,

\be
H^{'(1)}_{\a\b} = v^2 \,\e_{\a\b} \star_4 H^{(3)} \;, \;\;\;\;\; H^{'(2)}_\a = v^2 \e_{\a}{}^\b \star_4 H^{(2)}_\b\;, \;\;\;\;\; H^{'(3)} = \half \,v^2 \e^{\a\b} \star_4 H^{(1)}_{\a\b}
\ee
It follows that

\be
d \zeta' = v^2 \sqrt{\det G} \star_4 H^{(3)} \;, \;\;\;\;\; d \B'_\a = v^2 \, \e_\a{}^\b \star_4 d \B_\b + \zeta' \hat \e_{\a\b} d \A^\b + \frac{v^2 \zeta}{\sqrt{\det G}} \, G_{\a\b} \star_4 d \A^\b \label{beach}
\ee
These equations determine $\zeta'$ and $\B'_\a$, up to a constant shift. That these equations always have a solution - or, otherwise put, that $H^{(3)}$, $\A^\a$ and $\B_\a$ satisfy the necessary integrability conditions - is guaranteed by the consistency of the type IIB equations of motion. Now we perform an $STsTS$ transformation on the fields, using the formulae from the previous subsection, with

\be
\Sigma_2 = (1+ \hat \l_2 \zeta')^2 + \hat \l_2^2 \,v^4 e^{- 2 \phi} \det G
\ee
The  $STsTS$ transformation yields a new three-form field $F_3''$. The final RR three-form is related to it via

\be
F_3^f = \frac{\Sigma_2}{v^2} \star''_6 F_3''
\ee
After four T-dualities, the final metric is

\be
ds^2_f = \Sigma^{\half}_2 ds_4^2 +\S^{-\half}_2 ds_M^2 + \S^{-\half}_2 G_{\a\b} (dy^\a + \A^\a - \hat \l_2\, \hat \e^{\a\g} \B_\g')(dy^\b + \A^\b - \hat \l_2 \, \hat \e^{\b\g} \B'_\g)
\ee
and the components of the final $C^{(2)}$-field and the dilaton are given by

\be
d \zeta_{f} = d\zeta \;, \;\;\;\;\; e^{2\phi_f} = \S_2^{-1} e^{2\phi}
\ee

\be
H^{(2)}_{\a\, f} = (1+ \hat \l_2 \zeta')\, H^{(2)}_\a + \hat \l_2 v^2 e^{-2 \phi} \sqrt{\det G}\,  G_{\a\b} \star_4 d \A^{\b}
\ee
The last equation can be manipulated to yield

\be
d\B^f_\a = (1+ \hat \l_2\, \zeta')d\B_\a + \hat \l_2 \,\zeta\, v^2 \e_\a{}^\b \star_4 d\B_\b + \hat \l_2 v^2 \left(\zeta^2 + e^{-2\phi} \det G \right) \,\frac{G_{\a\b}}{\sqrt{\det G}} \star_4 d\A^\b
\ee
Finally, the expression for the new $\mathcal{C}_{\mu\nu}$ can be obtained from the new $H^{(3)}$

\be
H^{(3)}_f = \left[ (1+ \hat\l_2 \zeta')^2 - \hat \l_2^2 \, \left( e^{-2 \phi} v^4 \det G \right) \right] H^{(3)} + \frac{\hat \l_2 (1+\hat \l_2 \zeta')}{v^2 \sqrt{\det G}} \star_4 d \left( e^{-2 \phi} v^4 \det G \right)
\ee
Note that the internal $C$-field ($\zeta$) and internal metric (in  the S-dual NS-NS background) are invariant under the second generalized spectral flow, as is the four-dimensional Einstein metric. Therefore, we reconfirm that the second generalized spectral flow is nothing but the $\mathcal{S}$ transformation of the STU model\footnote{The four-dimensional description is probably more useful for obtaining the transformed $\mathcal{C}_{\mu\nu}$, as the $\mathcal{S}$ transformation acts as a fractional linear transformation on the four-dimensional axion-dilaton, the axion being the $4d$ Hodge dual of $\mathcal{C}_{\mu\nu}$.  }. 

One can obtain new six-dimensional supergravity solutions with two isometries by subsequently applying the generalized spectral flow transformations discussed above. In addition, one can perform a usual spectral flow, with parameter $\l_3$.
Given that the $\mathcal{S}$, $\mathcal{T}$ and $\mathcal{U}$ transformations commute, we conclude that the most general background that can be generated using spectral flows and their generalizations depends on three parameters, $\l_1, \l_2, \l_3$, and that, as conjectured in \cite{DGR}, the order in which the transformations are applied is irrelevant.

\subsection{Relationship to Melvin twists }

In the previous sections we have established that the generalized spectral flows of \cite{Bena:2008wt} correspond to a sequence of $STsTS$ and $T^4 S TsTST^4$ transformations, respectively. Realizing this fact makes it relatively easy to act with the given transformation on an arbitrary geometry with the required isometries, and the explicit form of the final result is given in the previous section.

As we have a CFT interpretation of the usual spectral flow diffeomorphisms \eqref{spflsptz} on the near-horizon geometry of the D1-D5 system, it would be very interesting to also understand the action of the two generalized spectral flows on the near-horizon geometry from a microscopic perspective. As it stands, the generalized spectral flow $STsTS$ has an interesting S-dual interpretation in terms of the CFT living on the worldsheet of a string propagating in the given background \cite{Frolov:2005dj}, but this is not a holographically dual description.

 The fact that the generalized spectral flow transformations involve an S-duality makes the microscopic interpretation rather difficult. Nevertheless, it is useful to note that the effect of the STsTS transformation on a type IIB background of the type \eqref{metfib} with purely RR three-form flux turned on is exactly equivalent to the following series of transformations

\bi
\item a T-duality along $y^1$
\item a lift to M-theory, using $ds_{11}^2 = e^{\frac{4 \phi}{3}} (dx^{11} + C^{(1)})^2 + e^{- \frac{2 \phi}{3}} ds_{10}^2$
\item a shift $y^2 \r y^2 + \hat \l \, x^{11}$
\item a dimensional reduction back to type IIA along the new coordinate $x^{11}$
\item a T-duality back on $y^1$
\ei
It is a matter of straightforward calculation to show that if one starts from the metric \eqref{metfib} with the $T^2$ directions labeled as $y^1$ and $y^2$, the final formulae we obtain from STsTS \eqref{tstmet} are identical to the expressions that we obtain from the T-Melvin twist -T procedure described above\footnote{For previous work in which the equivalence of the STsTS and T-Melvin-T transformations when acting on certain brane probes is implicitly present see \cite{ganpuff,aspuff}.}. The details of the computation are given in appendix \ref{dualid}.  Thus, we have an alternative description of the generalized spectral flows as T-Melvin-T and $T^5$-Melvin-$T^5$ transformations, respectively. We discuss several implications of this observation for the desired microscopic description in section \ref{micfeat}. 

Note that the Melvin twists dual to generalized spectral flows are those that involve the Taub-NUT fiber only, and from the perspective of the $\IR^4$ at the tip of Taub-NUT these correspond to equal twists in the two $\IR^2$s inside $\IR^4$. It has been known for quite some time that such equal-angle Melvin twists preserve a certain amount of the supersymmetry \cite{Gutperle:2001mb,FigueroaO'Farrill:2002tb}, and that they generically destroy the asymptotic structure of the spacetime. However, this is remedied by embedding the solution in Taub-NUT, and performing the Melvin twist along the Taub-NUT fiber, which stays finite at infinity.

\section{Spectral flows of the D1-D5-p-KK system \label{spfld1d5}}

In this section we apply the formulae for generalized spectral flows  derived in the previous section to the D1-D5-p-KK BPS and almost-BPS solutions. While this computation has been already done starting from much more general multicenter BPS and almost-BPS seed solutions, we find it useful to have the explicit formulae written down for this simple example directly in type IIB frame. A second motivation for performing this calculation is to exemplify the relative ease with which we can perform two generalized spectral flows on a given non-BPS solution.
As we have already discussed, the generalized spectral flows produce asymptotically Taub-NUT geometries which have a NHEK near-horizon.

\subsection{Review of the D1-D5-p-KK  solution}

The low-energy action for type IIB string theory in string frame is \cite{excl}

\be
S_{IIB} = \frac{1}{(2\pi)^7 \a'{}^4  } \int d^{10} x \sqrt{g} \left[ e^{-2\phi} \left(R +4 \p_\mu \phi \p^\mu \phi\right)
- \frac{1}{12} F^2 \right]
\ee
where $F$ is the RR three-form field strength. The extremal D1-D5-p-KK solution is

\be
ds^2=  \frac{H_0 R^2}{\sqrt{H_1 H_5}} \left( dy + (1-\frac{1}{H_0})
\frac{dt}{R}\right)^2 - \frac{  dt^2 }{H_0 \sqrt{H_1 H_5}} +
\sqrt{\frac{ H_1}{H_5}} \,ds^2_4 +
  \sqrt{H_1 H_5}\, ds^2_{HK} \label{metd15}
\ee
where $y \sim y + 2 \pi$ and the hyperk\"{a}hler metric $ds^2_{HK}$ is given by

\be
ds^2_{HK} = V (d\rho^2 + \rho^2 d \Om_2^2 ) + V^{-1} (d\psi + \cos \th d \phi)^2  \;, \;\;\;\;\;V = 1 + \frac{1}{\rho} \;, \;\;\;\;\; \rho = \frac{r^2}{4}
\ee
with

\be
e^{-2\phi} = \frac{H_5}{H_1} \;, \;\;\;\; H_5 = 1 + \frac{\a' g\, Q_5}{r^2} \;, \;\;\;\;\; H_1= 1 +
\frac{\a' g\, Q_1 v^{-1}}{r^2} \;, \;\;\;\;\;\; H_0 = 1 + \frac{\a'{}^2 g^2 n}{R^2 v\, r^2}
\ee
where $Q_1, Q_5$ and $n$ are the integer-quantized D1, D5 and momentum charges.
We will oftentimes use the shortcut

\be
H_i \equiv 1 + \frac{q_i}{r^2}= 1 + \frac{q_i}{4 \rho}
\ee
The internal space has total volume $(2\pi)^4 v \a'{}^2$. The RR three-form field strength is given by\footnote{The convention we use is $\e_{t r y \th \phi\psi}={ -}1$, which is the same as the one used in section \ref{adstonhek} upon noting  that the near-horizon time $\tau \propto -t$. }

\be
F= 2 q_5\, \varepsilon \,\om_{S^3} + \frac{2 g^2 \a' \e Q_1}{v}\, e^{-2 \phi} \star_6 \om_{S^3} = 2 q_5\,\varepsilon \, \om_{S^3}  {-}
\frac{2 q_1 R}{ r^3 H_1^2}\, dr \wedge dy \wedge dt
\ee
where $\om_{S^3}$ is the volume form of the unit three-sphere. Consequently, the two-form potential is

\be
C^{(2)} = - \frac{q_5}{4} \, \varepsilon \,\cos \th\, d\phi \wedge d\psi {-} (H_1^{-1} -1) R\, dy \wedge dt
\ee
The charges of the solution are

\be
Q_1 = \frac{v}{4 \pi^2 g \a'} \int e^{2\phi} \star_6 F \;, \;\;\;\;\;
Q_5 = \frac{1}{4 \pi^2 g \a'} \int F
\ee

\bigskip

\bigskip

\noindent If the system had not been embedded in Taub-NUT, i.e. if the function $V$ had been $V = \rho^{-1}$, the introduction of the parameter $\varepsilon = \pm 1$ would have had no effect, as the respective sign can always be switched by an interchange of $\phi$ and $\psi$. Otherwise put, the D1-D5-p solutions with either $\varepsilon = \pm 1$ are supersymmetric. This is no longer true once we embed the system in Taub-NUT, as the angles $\phi$ and $\psi$ are no longer interchangeable in the full solution, and the symmetry between $\varepsilon = 1$ and $\varepsilon = -1$ is broken. The solution with $\varepsilon = 1$ is BPS, whereas the one with $\varepsilon = - 1$ is  non-supersymmetric, of the type known as almost-BPS.

Before we turn on the Kaluza-Klein charge, the full D1-D5-p brane solution preserves four supersymmetries, in either the $T^4$ or K3 compactification. The supersymmetries are enhanced to eight in the near-horizon region, where the supersymmetry is of superconformal $(4,0)$ type. Adding one Kaluza-Klein charge does not change at all the near-horizon geometry, and thus the supersymmetry in the near horizon. Nevertheless, depending on whether the Taub-NUT solution preserves the $SU(2)_L$ or $SU(2)_R$,
the full geometry will either have four supersymmetries (enhanced to eight in the near horizon) or none (likewise enhanced to eight in the near horizon). All these statements only apply to the undeformed geometry which constitutes our departure point.

Given that upon embedding into Taub-NUT space the seed solutions with $\varepsilon = \pm 1$ become physically distinct, it becomes extremely confusing to keep using our conventions from section \ref{adstonhek}, in which a change in the sign of $\varepsilon$ was to be accompanied by an interchange\footnote{In those conventions, $H = \star H$ for both the BPS and almost-BPS solutions, and it is the self-duality condition necessary for preserving supersymmetry that differs.} of $\psi$ with $\phi$. Therefore in this section we do not perform the interchange, and thus the two solutions are distinguished by $H = \varepsilon \star H$. We comment at the end of this chapter on the additional sign changes one needs to perform in order to match to section \ref{adstonhek}.

Next, we consider generalized spectral flows of the full D1-D5-p-KK geometry. The TsT transformations involved have to 
respect the $SU(2)$ symmetry not broken by the Kaluza-Klein monopole. Depending on whether the original background was supersymmetric or not, the TsT transformations will either preserve supersymmetry (if $\varepsilon=1$), yielding a generic five-dimensional supersymmetric solution with a Taub-NUT base, of the type discussed in \cite{Gauntlett:2004qy,Bena:2005ni,Bates:2003vx}, or break supersymmetry (when $\varepsilon =-1$), this time generating the rather complicated backgrounds of \cite{DGR}.

To simplify our formulae, in the following we set $\a' = g =1$.

\subsection{The effect of spectral flows}

In order to apply our formulae for spectral flow, we need to write the background we start from in the form \eqref{metfib} - \eqref{decc}. Our input is then

\be
G_{\a\b} = \left(\begin{array}{cc}  \frac{H_0 R^2}{\sqrt{H_1 H_5}} & 0 \\ 0 &  \frac{4 \sqrt{H_1 H_5}}{V} \end{array} \right)\;, \;\;\;\;\;\;  e^{2 \phi} = \frac{H_1}{H_5}
\ee
where the $T^2$ coordinates $y^\a$ are taken to be $y$ and $ \half \psi$, both of which have period $2\pi$. The starting four-dimensional gauge fields are

\be
\A^\a = \left(\begin{array}{c}  \vspace{3mm} (1-H_0^{-1}) R^{-1} \,dt \\ \half \cos\th d\phi \end{array} \right)\;, \;\;\;\;\;\; \B_\a =\left(\begin{array}{c}  \vspace{3mm}  { -}R (1-H_1^{-1})\, dt \\  - \frac{q_5}{2} \, \varepsilon \,\cos \th\, d\phi \end{array} \right) \label{stgf}
\ee
and the eight-dimensional base metric is

\be
ds_8^2 = ds_4^2 + ds^2_M
\ee
with
\be
ds^2_4 =    - \frac{  dt^2 }{H_0 \sqrt{H_1 H_5}}  + \sqrt{H_1 H_5} V (d \rho^2 + \rho^2 d \Om_2^2)\;, \;\;\;\;\; ds^2_M = \sqrt{\frac{ H_1}{H_5}} \sum_{i=1}^4 dx_i^2
\ee
Additionally, the starting background has $\zeta = \mathcal{C}_{\mu\nu} =0$. The $\Sigma$ factor entering the first spectral flow (with parameter $\l_1$) is

\be
\Sigma_1 = 1 + \frac{\l^2_1 H_0 H_5 R^2}{H_1 V} \equiv 1 + \l_1^2 X
\ee
The spectral-flowed metric is

\bea
ds^2 & = & \frac{H_0 R^2}{\sqrt{H_1 H_5 \S_1}} \left( dy + \frac{dt}{R} (1-H_0^{-1}) + \frac{\l_1  q_5}{4} \varepsilon \cos \th d\phi \right)^2 +\non \\
& + &  \frac{\sqrt{H_1 H_5}}{V \sqrt{\S_1}} \left( d \psi + \cos \th d\phi { -}\l_1 R (1- H_1^{-1}) dt\right)^2 + \sqrt{\frac{H_1 \S_1}{H_5}} ds_M^2 + \non \\
&+& \sqrt{H_1 H_5 \S_1} V (d\rho^2 + \rho^2 d\Om_2^2) - \frac{\sqrt{\S_1}}{H_0 \sqrt{H_1 H_5}} \, dt^2
\eea
The new components of the $C^{(2)}$ field are determined from

\be
\zeta_1 = \frac{2 \l_1 X}{1 + \l_1^2 X} \;, \;\;\;\;\; \mathcal{C}_{\mu\nu}=0
\ee
using \eqref{decc} and \eqref{newparamone}. The new value of the dilaton is

\be
e^{2 \phi} = \frac{H_1 \S_1}{H_5}
\ee
For $\varepsilon =1$, this is just the BPS spectral flow, and our formulae agree with those of \cite{Bena:2008wt}. For $\varepsilon =-1$, we obtain an asymptotically Taub-NUT black hole whose near-horizon geometry is warped $AdS_3$.

Next, we proceed to performing a second spectral flow. For this, we need to solve for $\zeta'$ and $\B'_\a$ in \eqref{beach}. We already know that the first spectral flow does not change the value of $H^{(3)}$, and on the D1-D5 background this is simply zero. Consequently, $d \zeta' =0$, and we choose the simplest solution $\zeta'=0$.  The parameter $\Sigma_2$ now becomes

\be
\Sigma_2 = 1 + \frac{\l_2^2 H_1 H_0 R^2}{H_5 V} \equiv 1 + \l_2^2 \tilde{X}\ ,
\ee
and the equation for $\B'_\a$ simplifies to

\be
d \B'_\a = \frac{2 \l_1 H_0 R^2}{V} \frac{G_{\a\b}}{\sqrt{\det G}} \star_4 d\A^\b + \frac{H_1}{H_5} \e_\a{}^\b \star_4 d\B_\b
\ee
where the fields that enter are given in \eqref{stgf} (they are the gauge fields  \emph{before} the STsTS transformation). The solution is  

\be
\B'_y = {-}\frac{\l_1 R^2 q_0}{4} \cos \th d \phi {+} \frac{\varepsilon R}{ H_5} dt\;, \;\;\;\;\;\B'_{\hat \psi} =- \frac{q_1}{2} \cos \th d \phi {-}\frac{2 \l_1 R}{V} dt
\ee
The final metric is thus

\bea
ds^2 & = & \frac{H_0 R^2}{\sqrt{H_1 H_5 \S_1 \S_2}} \left( dy + \frac{dt}{R} (1-H_0^{-1}){+} \frac{\l_1  \l_2 R}{V} dt + \frac{\l_1  q_5 \, \varepsilon + \l_2 q_1}{4}  \cos \th \, d\phi \right)^2 +\non \\
& + &  \frac{\sqrt{H_1 H_5}}{V \sqrt{\S_1 \S_2}} \left( d \psi + \cos \th d\phi {-} \frac{\l_1\l_2 R^2 q_0}{4} \cos \th d\phi {-} \l_1 R dt {+}\frac{\l_1 R}{H_1}  dt {+} \frac{\l_2 R \varepsilon}{H_5} dt\right)^2 +  \non \\
&+&  \sqrt{\frac{H_1 \S_1}{H_5 \S_2}} ds_M^2 + \sqrt{H_1 H_5 \S_1 \S_2} V (d\rho^2 + \rho^2 d\Om_2^2) - \frac{\sqrt{\S_1 \S_2}}{H_0 \sqrt{H_1 H_5}} \, dt^2
\label{uglymetric}
\eea
and the final value of the dilaton

\be
e^{2\phi} = \frac{\S_1 H_1}{\S_2 H_5}= \frac{1+ \l_1^2 X}{1+ \l_2^2 \tilde{X}} \frac{H_1}{H_5}
\ee
The component of the final $C^{(2)}$ field satisfies

\be
d \B_\a^f = d \B_\a + \frac{2 \l_2 H_0 R^2}{V} \frac{G_{\a\b}}{\sqrt{\det G}} \star_4 d \A^\b
\ee
which yields

\be
\B_y^f = {-}R (1- H_1^{-1}) dt {-} \frac{\l_2 q_0 R^2}{4} \cos \th d\phi \;, \;\;\;\;\; \B^f_{\hat \psi} = - \frac{q_5 }{2} \varepsilon \cos \th d\phi {-} \frac{2  \l_2 R dt}{V}
\ee
We also list the expression for the final $H^{(3)}$

\bea
H^{(3)}_f & = & \frac{2 \l_2}{v^2 \sqrt{\det G}} \star_4 d \tilde{X} = {-}\frac{\l_2 H_5 V}{H_1 H_0 R} \rho^2 \tilde{X}' \sin \th dt \wedge d\th \wedge d\phi \non \\
& = &{-} \l_2 R  \left(\frac{q_5}{4 H_5} + \frac{1}{V} - \frac{q_0}{4 H_0} - \frac{q_1}{4 H_1}\right)
 \sin \th dt \wedge d\th \wedge d\phi
\eea
from which one can easily determine the final $\mathcal{C}_{\mu\nu}^f$ component of the $C^{(2)}$ field, using \eqref{defh3}. Nevertheless, since these particular components do not enter the computation of the physical charges, we do not include herein the corresponding formulae.

It is interesting to observe the very small difference between the solutions  (\ref{uglymetric}) coming from the spectral flow of BPS and almost-BPS geometries. In contrast, when expressing the solutions in the duality frame in which the D1, D5 and P charges correspond to three sets of M2 branes wrapping $T^2$s inside $T^6$, the solution of \cite{DGR} looks nothing like a BPS solution. Our calculation shows that this difference is only apparent, and comes from the choice in writing the metric as a sum of squares in which one factorizes first the $dt$ fiber and then the $d \psi $ fiber. If one choses to write the BPS and the spectral-flowed almost-BPS solutions in the manner of (\ref{uglymetric}), their similarity is striking.

\subsection{Properties of the solution}

As advertised at the beggining of this section, performing two generalized spectral flows on the almost BPS D1-D5-p-KK solution produces a geometry that interpolates between $\mathbb{R} \times TN \times S^1$ asymptotically and a spacetime locally equivalent to six-dimensional NHEK in the near horizon. Indeed, the near horizon limit of \eqref{uglymetric} matches perfectly with our formulae from section \ref{adstonhek}, provided we identify 

\be
c = \hat \l_1=  \half \l_1 \;, \;\;\;\;\; \tilde{c} = - \hat \l_2 =- \half \l_2 \label{sgndiff}
\ee
where we have used the fact that for TsT $d=1$ and

\be
\V = \sqrt{\frac{q_0 q_5}{q_1}}  {R} \;, \;\;\;\;\;\;\;\;\tilde \V = \sqrt{\frac{q_0 q_1}{q_5}}  {R}
\ee
The sign difference in \eqref{sgndiff} can be explained by our different conventions in this section, namely by the fact that we have chosen to not interchange $\psi$ and $\phi$ when $\varepsilon=-1$. This effectively amounts to having $H \r - \star H$ under the Hodge dualities required by the two electromagnetic transformations, which in turn  implies that in our original conventions it is $- H^{(n)}$ which undergoes the transformations \eqref{trh}. Finally, this amounts to changing the sign of $\l_2$.

\subsubsection*{Conserved charges}

We would also like to compute the conserved charges of the solution, whose expressions are

\be
Q_5 = \frac{1}{4 \pi^2} \int_{\S_3}  F^f \;, \;\;\;\;\; Q_1 = \frac{1}{4 \pi^2} \int_{\S_3} e^{2\phi_f}  \star F^f
\ee
The closed surface $\S_3$ can be either $S^3$ or $S^2 \times S^1_{y}$, thus yielding the usual and dipole charges of each kind.
The only components of the final $F$-field that we need to compute are the $F^f_\a = \half F^f_{\mu\nu\a} \, dx^\mu \wedge dx^\nu$ ones, given b

\be
F_{\a} = d \B_\a - \zeta \hat \e_{\a\b} d\A^\b + \hat \e_{\a\b} \A^\b \wedge d\zeta  = H^{(2)}_{\a} - H_{\a\b}^{(1)} \wedge \A^\b
\ee
where we have dropped the superscript `$f$' from all fields. We also need the components of $(\star F)_\a$, which in light of the above expression are simply given by

\be
(\star F)_\a = \e_a{}^\b \star_4 H^{(2)}_\b - \e_{\a\b} (\star_4 H^{(3)}) \wedge \A^\b
\ee
The associated charges are 
\be
\mathcal{Q}_{5} = \frac{q_5 \varepsilon - \l_1  \l_2 q_1 R^2}{1+ \l_1^2 R^2}\, \varepsilon \;, \;\;\;\;\; \mathcal{Q}_{5;dip} = \frac{4 \l_1 R^2 {+} \l_2 q_0 R^2 }{1+ \l_1^2 R^2} \, \varepsilon
\ee

\be
\mathcal{Q}_{1} = \frac{q_1  - \l_1  \l_2 q_5 \varepsilon R^2}{1+  \l_2^2 R^2} \;, \;\;\;\;\; \mathcal{Q}_{1;dip} = \frac{4  \l_2 R^2 {+ } \l_1 q_0 R^2 }{1+ \l_2^2 R^2}
\ee
where the extra factor of $\varepsilon$ accounts for interchanging $\phi$ and $\psi$. These charges have been computed before we perform the final $\mathcal{U}$ transformation. Note that when $q_1=q_5,\,\lambda_1=-\lambda_2$, the solution (\ref{uglymetric}) is related to the solution of \cite{stromwei} by a constant rescaling on both the metric and the gauge field. The solution also carries nontrivial Taub-NUT charge, which we do not compute. In \cite{stromwei}, the Taub-NUT charge was fixed to one.

Under the final $\mathcal{U}$ transformation, the charges transform as

\be
\left(\begin{array}{c} \mathcal{Q}_5^f \\  \mathcal{Q}_{5; dip}^f \end{array}\right) \r  \s_3 \, \mathcal{U}\, \s_3
\left(\begin{array}{c}   \mathcal{Q}_5 \\  \mathcal{Q}_{5; dip} \end{array}\right)
\ee
and similarily for $ \mathcal{Q}_1$, where the matrix $\mathcal{U}$ of the transformation is given by \eqref{finu}.

\section{An infinite family of NHEK embeddings in string theory \label{infam}}

Having shown that one can start from a D1-D5-p black hole embedded in Taub-NUT in a non-supersymmetric almost-BPS fashion and obtain a spacetime with a NHEK geometry in the infrared by two generalized spectral flows, it is clear that there should exist a much larger family of solutions that have such an infrared geometry.
Indeed, the NHEK geometry is obtained by transforming the near-horizon geometry of the original D1-D5-p solution, and it has been known for a long time \cite{Ferrara:1995ih,Strominger:1996kf,Ferrara:1996dd,Sen:2005wa} that this near-horizon geometry does not change if one changes the moduli of the solution as long as the charges remain the same.

Thus, by starting with an asymptotically-Taub-NUT  D1-D5-p solution with arbitrary moduli and performing two generalized spectral flows with the coefficients above, one can obtain an asymptotically Taub-NUT solution with a NHEK infrared. The example worked out in the previous section corresponds to a subclass of this infinite family, in which the only moduli that are allowed to vary at infinity are the sizes of various cycles, and correspond to the constants in the D1, D5, p and Taub-NUT harmonic functions. However, it is clear that there should exist solutions that also have nontrivial axions and have a NHEK infrared.

The idea is to start from the most general three-charge black hole geometry embedded non-supersymmetrically in Taub-NUT, which belongs to the so-called almost-BPS class of solutions found in \cite{Goldstein:2008fq,Bena:2009ev}. This black hole is a generalization of the one considered in the previous section in that various moduli (axions) are turned on at infinity. The effect of three (generalized) spectral flows on arbitrary almost-BPS backgrounds has been worked out in \cite{DGR} and, rather than repeating the computation of section \ref{spfld1d5} with this more general input data, we can simply recover the final answer by  setting certain parameters of the solution found in \cite{DGR} to their appropriate values.

To relate the formalism and notation of \cite{DGR} to that used in this paper we first extract the solution corresponding to the generalized spectral flows of a D1-D5-p-TN geometry with no axions, and relate it to the solution we obtained in the previous section. We then show how to obtain the most general geometry with a NHEK infrared.

\subsection{The solution with no axions}

An almost-BPS solution with no axions is determined by five harmonic functions: $Z_1$, $Z_2$, $Z_3$, $V$ and $M$, corresponding respectively to D1, D5, p, KK-Monopole and KK-momentum charges.  In the eleven-dimensional duality frame in which the D1, D5, and p charges correspond to three different species of M2 branes wrapping three orthogonal $T^2$'s inside a $T^6$, the metric and fields of the solution both before and after the generalized spectral flows are given by equations (6.6) and (6.7) of \cite{DGR}, with the implicit understanding that un-tilded quantities describe the solution before the spectral flows and tilded ones the one after.

It is rather straightforward to dualize the solution before the spectral flow to the D1-D5-P frame solution by
dimensionally-reducing along one of the torus directions,
and then performing three T-dualities, as explained in detail in \cite{Bena:2008dw}. Since the solution has neither axions nor rotation, and since the four-dimensional base space is Taub-NUT,  the metric is simply
\bea
ds^2&=& -{1\over \sqrt{Z_1 Z_2} Z_3}(dt)^2 + {Z_3\over \sqrt{Z_1 Z_2}} (dz+A^3)^2 \non \\
&+& \sqrt{Z_1 Z_2}  \left(V^{-1} (d \psi + A)^2 + V ds^2_{\IR_3}\right) + \sqrt{Z_1\over Z_2} \sum_{i=1}^4 dx_i^2 \, 
\label{d1d5p}
\eea
where the coordinate $z$ is related to the coordinate $y$ we used previously by $z= R \, y + t$.
The functions $Z_1,Z_2,Z_3$ and $V$ are harmonic in
$\IR^3$, the forms $A$ and $A^I$ are given by
\be
 A^I=-{dt \over
  Z_I} ~,~~~ d A = *_{\IR_3} d V
\ee
and the relation with the harmonic functions we used in section \ref{spfld1d5} is quite clear:
\be
H_1=Z_1~,~~~H_5=Z_2~,~~~ H_0=Z_3~,~~~ V=V.
\ee
Also, since we start from a non-rotating solution, we have taken $M=0$. If the original solution has no axions, then the other three harmonic functions $K_1,K_2$ and $K_3$ are also zero. The two generalized spectral flows we perform correspond to $\gamma_1$ and $\gamma_2$ spectral flows. In the M2-M2-M2 duality frame the most general solution after three spectral flows is given in Section 6 of \cite{DGR}. This solution can be dualized to the D1-D5-P duality frame, and its metric and dilaton are:

\bea
ds^2_{IIB} &=& - \ds \frac{1}{\widetilde Z_3\sqrt{\widetilde Z_1\widetilde Z_2}}(dt+\widetilde \mu (d \psi + \widetilde A))^2  + \ds\frac{\widetilde Z_3}{\sqrt{\widetilde Z_1 \widetilde Z_2}}(dz+\widetilde A^3)^2  \nonumber \\
&+& \sqrt{\widetilde Z_1 \widetilde Z_2} \left(\widetilde V^{-1} (d
  \psi + \widetilde A)^2 + \widetilde V ds^2_{\IR_3}\right) +
\ds\sqrt{\frac{\widetilde Z_1}{\widetilde Z_2}} \sum_i dx_i^2    \label{newD1D5Pmetric} \\ e^{
2\phi} &=&  \ds\frac{\widetilde
    Z_1}{\widetilde Z_2},  \nonumber \eea
where we have used the fact that the original solution had no rotation when compactified to 3+1 dimensions. The NS-NS B-field is zero and the RR three-form field strength can also be obtained by a straightforward if tedious exercise from \cite{DGR}.

The one-form $\widetilde A^3$ is:
\be
\widetilde A^3 = -{ 1 \over \widetilde{W}_3 } \left( dt +\widetilde{\mu}(d\psi+\widetilde{A})  \right) + \widetilde{P}_3 \, (d\psi + \widetilde{A})+ \widetilde{w}^3
\ee
and the forms $\widetilde A^1$ and $\widetilde A^2$, which only appear
in the expression for the RR three-form, have similar expressions. The
resulting solution after three spectral flows on an arbitrary
almost-BPS solution can be found in \cite{DGR}.

Since in the previous section we have performed the two spectral flows
directly in six-dimensional supergravity, and in \cite{DGR} they were
performed by a combination of six T-dualities on the compactification six-torus and three gauge
transformations, it is instructive to confirm that the final result is
the same. The parameters of the new solution, in the language of section 6 of \cite{DGR}, are:

\be
T_I=1~,~~ N_1 = \gamma_1^2 Z_2 Z_3 + V Z_1~,~~N_2 = \gamma_2^2 Z_1 Z_3 + V Z_2~,~~N_3=V Z_3
\ee
\be
\widetilde V = V  {-} \gamma_1 \gamma_2 Z_3 ~,~~~ \widetilde \mu =  - {V Z_3 \over \widetilde V^2}(\gamma_1 Z_2 + \gamma_2 Z_1)
\ee
\be
 \widetilde{A} = A  {-}  \gamma_1 \gamma_2 \, v_3 ~,~~~{\rm with}~~~ \star_{\IR_3} d v_I =  - d Z_I
\ee
The three warp factors are
\be
\widetilde Z_1 = { \gamma_1^2 Z_2 Z_3 + V Z_1 \over V  {-} \gamma_1 \gamma_2 Z_3}~,~~
\widetilde Z_2 = { \gamma_2^2 Z_1 Z_3 + V Z_2 \over V  {-} \gamma_1 \gamma_2 Z_3}~,~~
\widetilde Z_3 = {V Z_3 \over V  {-} \gamma_1 \gamma_2 Z_3}
\ee
the three scalar functions giving the electric field are
\be
\widetilde W_1 = { \gamma_1^2 Z_2 Z_3 + V Z_1 \over V - \gamma_1 \gamma_2 Z_3 }~  {= \widetilde Z_1~},~~ \widetilde W_2 = { \gamma_2^2 Z_1 Z_3 + V Z_2 \over V- \gamma_1 \gamma_2 Z_3}~ {= \widetilde Z_2~},~~ \widetilde W_3 = {V Z_3 \over V + \gamma_1 \gamma_2 Z_3} .
\label{floating}
\ee
The three scalars giving the components of the magnetic fields along the $\psi$ direction are:
\be
 \widetilde P_1= {\gamma_1 Z_2 Z_3 \over N_1}+{\widetilde \mu \over \widetilde W_1}~,~~  \widetilde P_2= {\gamma_2 Z_1 Z_3 \over N_1}+{\widetilde \mu \over \widetilde W_2}~,~~  \widetilde P_3= {\widetilde \mu \over \widetilde W_3}
\ee
and the three vectors giving the $\IR^3$ component of the magnetic fields are
\be
 \widetilde{w}^1 =  \gamma_2 \, v_3 ~,~~~
 \widetilde{w}^2 =  \gamma_1\, v_3 ~,~~~
 \widetilde{w}^3 =  \gamma_1 \, v_2   {+} \gamma_2 \, v_1~.
 \ee
The six-dimensional part of the metric ({\ref{newD1D5Pmetric}}) can then be simplified to 

\bea
ds^2_{IIB} &=& \ds \frac{- (dt+\widetilde \mu (d \psi + \widetilde A))^2}{\widetilde Z_3\sqrt{\widetilde Z_1\widetilde Z_2}}
+{\widetilde Z_3 \over \sqrt{\widetilde Z_1\widetilde Z_2}} \left(dz - {dt \over \widetilde W_3 }  {+} \gamma_1 \, v_2  + \gamma_2 \, v_1 \right)^2  \nonumber \\
&+& \sqrt{\widetilde Z_1 \widetilde Z_2} \left(\widetilde V^{-1} (d \psi + \widetilde A)^2 + \widetilde V ds^2_{\IR_3}\right)
\label{uglymetric2}
\eea
which one can check straightforwardly that is the same as the metric obtained after two spectral flows in the previous section in equation (\ref{uglymetric}). To see this one should use the fact that $\gamma_1 =\lambda_1 R $ and $\gamma_2 = - \lambda_2 R$ and make use of rather nontrivial relations such as
\be
\widetilde Z_1 \widetilde Z_2  - {\widetilde \mu^2  \widetilde V \over  \widetilde Z_3}= Z_1 Z_2   ~.
\ee

The main reason why the two metrics (\ref{uglymetric}) and (\ref{uglymetric2}) look so different is that the fibers are completed in different orders: in (\ref{uglymetric}) the $y$ and $\psi$ fibers are completed before the $t$ fiber, while in (\ref{uglymetric2}) the $t$ fiber is completed after the $z (=y)$ fiber but before the $\psi$ fiber, as it is common for supersymmetric solutions.

\subsection{A more general solution}

Having shown how to obtain the solution of section \ref{spfld1d5} from the class of solutions found in \cite{DGR}, we now write down a much larger family of solutions that have a NHEK infrared.

As it is well known \cite{Goldstein:2008fq,Bena:2009ev}, if one turns on nontrivial Wilson lines in an almost-BPS solution, the metric is not affected at all, and neither are any of the field strengths. From a four-dimensional perspective, these Wilson lines corresponds to some combination of axions, and in the language of the harmonic functions used to construct such solutions, they is given by a constant in the $K_I$.

Performing two generalized spectral flow transformations\footnote{And a $\mathcal{U}$ coordinate transformation, which we will neglect in this section.} on an almost-BPS solution with axions turned on, we find a geometry whose infrared is still NHEK. However, the full asymptotically-flat solution is changed, and depends non-trivially on the $K_I$. Hence one can obtain a much larger family of NHEK embeddings into string theory, determined a-priori by three extra continuous parameters.

The solution is a straightforward modification of the one in the previous subsection
\be
T_I=1+\gamma_I K_I ~,~~ N_1 = \gamma_1^2 Z_2 Z_3 + V T_1^2 Z_1~,~~N_2 = \gamma_2^2 Z_1 Z_3 + V T_2^2 Z_2~,~~N_3=V Z_3 T_3^2
\ee
\be
\widetilde V = T^3 V - \gamma_1 \gamma_2 T_3 Z_3 ~,~~~ \widetilde \mu =  - {V T_3 Z_3 \over \widetilde V^2}(\gamma_1 T_2 Z_2 + \gamma_2 T_2 Z_1)
\ee
\be
 \widetilde{A} = A -  \gamma_1 \gamma_2 \, v_3 ~,~~~{\rm with}~~~ \star_{\IR_3} d v_I =  - d Z_I
\ee
where $T^3 \equiv T_1 T_2 T_3$ and we have kept the dependence on $T_3$ implicit, despite the fact that when $K_3$ is finite and $\gamma_3 = 0 $, $T_3$ is equal to one. The reason is that neither the metric, nor the fields of the original solution depend on $K_3$, and one should therefore be able to obtain different solutions with an NHEK infrared region by taking the limits $K_3 \rightarrow \infty~,~\gamma_3 \rightarrow 0$ while keeping $\gamma_3 K_3$ fixed. 
The three warp factors are

\be
\widetilde Z_1 = { \gamma_1^2 Z_2 Z_3 + V T_1^2 Z_1 \over T^3 V - \gamma_1 \gamma_2 T_3 Z_3}
~,~~ \widetilde Z_2 = { \gamma_2^2 Z_1 Z_3 + V T_2^2  {Z_2} \over T^3 V - \gamma_1 \gamma_2 T_3 Z_3}
~,~~ \widetilde Z_3 = {V Z_3 T_3^2 \over T^3 V - \gamma_1 \gamma_2 T_3 Z_3}
\ee
the three scalar functions giving the electric field are
\be
\widetilde W_1 = { \gamma_1^2 Z_2 Z_3 + V T_1^2 Z_1 \over T^3 V  {-} \gamma_1 \gamma_2 T_3 Z_3}  {= \widetilde Z_1}~,~~
\widetilde W_2 = { \gamma_2^2 Z_1 Z_3 + V T_2^2 Z_2 \over T^3 V  {-}\gamma_1 \gamma_2 T_3 Z_3}  {= \widetilde Z_2} \non
\ee
\be
\widetilde W_3 = {V Z_3 T_3^2 \over T^3 V  {+} \gamma_1 \gamma_2 Z_3} \neq \widetilde Z_3~.
\label{floating2}
\ee
The three scalars giving the components of the magnetic fields along the $\psi$ direction are now
\be
 \widetilde P_1= {\gamma_1 Z_2 Z_3  {+ V Z_1 T_1 K_1 }\over N_1}+{\widetilde \mu \over \widetilde W_1}~,~~  \widetilde P_2= {\gamma_2 Z_1 Z_3  {+ V Z_2 T_2 K_2}\over N_1}+{\widetilde \mu \over \widetilde W_2}~,~~  \widetilde P_3= {\widetilde \mu \over \widetilde W_3}~,
\ee
and  the three vectors giving the $\IR^3$ component of the magnetic fields are unchanged
\be
 \widetilde{w}^1 = \gamma_2 \, v_3 ~,~~~
 \widetilde{w}^2 =  \gamma_1\, v_3 ~,~~~
 \widetilde{w}^3 =  \gamma_1 \, v_2   {+} \gamma_2 \, v_1~.
 \ee
The six-dimensional part of the metric ({\ref{newD1D5Pmetric}}) can again be simplified to
\bea
ds^2_{IIB} &=& \ds \frac{- (dt+\widetilde \mu (d \psi + \widetilde A))^2}{\widetilde Z_3\sqrt{\widetilde Z_1\widetilde Z_2}}
+{\widetilde Z_3 \over \sqrt{\widetilde Z_1\widetilde Z_2}} \left(dz - {dt \over \widetilde W_3 }  {+} \gamma_1 \, v_2  + \gamma_2 \, v_1 \right)^2  \nonumber \\
&+& \sqrt{\widetilde Z_1 \widetilde Z_2} \left(\widetilde V^{-1} (d \psi + \widetilde A)^2 + \widetilde V ds^2_{\IR_3}\right)
\label{uglymetric2a}
\eea
It is interesting to observe in (\ref{floating}) that the
solutions obtained after two generalized spectral flows still have two
of the electric field functions equal to the corresponding warp
factors, and hence they still admit  ``floating'' D1 and D5 branes, that can be
placed anywhere in the solution without feeling a force. This is visible in equation. (\ref{floating}) from the fact that
$\widetilde W_1 = \widetilde Z_1$ and $\widetilde W_2 =
\widetilde Z_2$.

The most general solution of minimal $6d$ supergravity with a NHEK infrared should be a slight generalization of the above one, in that one can allow $\g_3$ and $K_3$ to be finite and also  $M \neq 0$. All the necessary formulae can be found in section 6 of \cite{DGR}. While the most general solution appears to be parameterized by the eight constants in the harmonic functions, two of these parameters are redundant, as the moduli space of the STU model is just six-dimensional.

\section{Microscopic features of NHEK spacetimes \label{micfeat}}

As mentioned in the Introduction, the main purpose of our endeavor is to obtain a microscopic theory dual to Kerr black holes. Our construction opens two different routes towards this:

The first is to explore the effect of our pseudo-duality transformations on the D1-D5 field theory, and to attempt to construct a theory that is a deformation of the D1-D5 field theory and  flows in the infrared to the CFT dual to NHEK. Indeed, the effect of a single Melvin twist on the super Yang-Mills theory living on the worldvolume of a stack of D4 branes is to deform it by a hypermultiplet mass term proportional to the strength of the Melvin twist \cite{chocky}. It is reasonable then to expect that the effect of a combination of T-dualities and Melvin twists on a stack of D5 branes is likewise to add a  relevant operator to the six-dimensional Super-Yang-Mills theory living on this stack of branes. Similarly, the effect of 5 T-dualities, a Melvin twist and 5 T-dualities back will deform the 1+1 dimensional Super-Yang-Mills theory living on a stack of D1 branes by a certain  operator. The effect of the T-Melvin-T transformation on D1 branes may perhaps be incorporated by studying instantons in the deformed D5-brane gauge theory, or possibly by incorporating more exotic effects along the lines of \cite{ganpuff,aspuff}. Similarly, one can try to find the effect of the pseudo-dualities transformations on the 1-5 strings. The net result of this analysis (which is achievable, but lies beyond the scope of the present paper) should be an explicit deformation of the Lagrangian of the D1-D5 field theory, which redirects the RG flow in the infrared from the usual D1-D5 CFT to the CFT dual to NHEK.

Another route to the microscopic theory is to use the infinite family of brane embeddings found in the previous section. Clearly the physics of these branes is described in the regime of parameters where the branes do not backreact on the the geometry by a certain theory (most likely a gauge theory), and this theory should flow in the infrared to the CFT dual to NHEK. However, given a certain non-supersymmetric combination of many types of branes in many types of background electric and magnetic fields, reading off this theory is not easy. The advantage of having a large family of brane configurations, like the ones we have explicitly constructed, is that one can focus on particular ones (where for example the background magnetic fields after the dualities are zero) for which reading off the corresponding field theory is much easier than for the general configuration. Finding these simple brane configurations in the haystack of solutions constructed in Section \ref{infam} is quite tedious, and we leave it to future work.

Even though we lack so far a theory dual to the Kerr black hole, we can nevertheless use our construction to predict from the dual bulk solutions certain features that this theory will have to corroborate. In particular, we will show that this theory will have a moduli space of deformations parameterized by at least  several continuous parameters. These deformations will change the topology but not the asymptotics of the NHEK spacetime, and hence will not be visible in perturbation theory. Understanding the structure of the space of asymptotically-NHEK solutions is also likely to reveal many features of the dual theory, much like it happens for example in the D1-D5 CFT \cite{Lunin:2001fv,Lunin:2002iz}.

What interests us in particular is to understand what happens to the
Coulomb and Higgs branches of the original D1-D5 gauge theory under generalized spectral flow, and how to characterize the moduli space of deformations around the geometries with an NHEK IR.

\subsection{Spectral flows of the Coulomb branch}

In this subsection we study the effect of two generalized spectral flows on various configurations
on the Coulomb branch of the D1-D5 gauge theory, and find the objects parameterizing the corresponding
branch of the moduli space in the theory dual to NHEK. One can easily use the machinery of section \ref{infam} to write down the solutions for the most general configurations, and then proceed to analyze in detail these solutions. However, it is much more useful to develop first a physical intuition of what T-Melvin-T transformations do to the D1 and D5 branes that make up the original solution, and then to use this to understand the physics of the full solution.

\subsubsection*{The effect of a single Melvin twist/generalized spectral flow on D5 branes}

In order to understand the effect of Melvin twists on the microscopic
features of the solution it is useful to remember the action of Melvin
twists on a stack of D4 branes. When one spreads these D4 branes
uniformly on a circle, uplifts to M-theory, and reduces with a Melvin
twist on the circle on which the D4 branes are spread, the resulting
configuration is an NS5 brane wrapping this circle, with the D4 brane
charge dissolved in it. Only for certain Melvin twist strengths does
this NS5 brane have an integer dipole charge, and hence this
configuration is physical.

From a ten-dimensional perspective the Melvin twist induces some
nontrivial fields, and if one places a stack of D4 branes in these
fields they can polarize into a circular NS5 brane by the Myers effect
\cite{Myers:1999ps}. Normally, this circular NS5 brane would collapse, and
these fields prevent it from doing so.

Hence, what before the Melvin twist is a particular configuration on
the Coulomb branch of the D4 branes, becomes after the Melvin twist a
configuration on a nontrivial branch of the moduli space of the new
theory \cite{chocky}. Note that other configurations on the original Coulomb branch
do not survive the Melvin twist, so only certain Coulomb branch
configurations will map to the moduli space of the new theory.
Moreover, it is likely that these configurations do not map to the
entire moduli space of the new theory, but only to a small portion
thereof.

Note that the configurations we find have an NS5 brane dipole charge,
and hence are intrinsically nonperturbative: they are not visible in
the Lagrangian of the field theory living on the D4 branes, and, as explained in
 \cite{chocky2}, their existence can only be inferred using integrability \cite{Nekrasov:1996cz} or the Dijkgraaf-Vafa relation between
five-dimensional theories and Matrix Quantum Mechanics \cite{Hollowood:2003gr, Dijkgraaf:2002dh}.
If one dualizes this configuration back to
the D1-D5 frame, the D4 brane becomes a D5 brane, and the NS5 dipole
brane becomes a KK-monopole whose special direction is along the
common D1-D5 direction. Hence, after the T-Melvin-T transformation the
Coulomb branch of the D5 branes is mapped into a nonperturbative branch
of the moduli space, in which the D5 branes polarize into a KKM wrapping an $S^1$
inside the space transverse to the branes.

Now, as we have shown in Section 4, the T-Melvin-T transformation
described above is the same as a generalized spectral flow with
parameter $\gamma_1$. Hence, one can also translate the mapping of the
moduli spaces described above in the language of multicenter solutions
and spectral flows thereof.
The solution describing D5 branes in $\IR^4$ can be embedded in
Taub-NUT, where it becomes a solution describing D5 branes at the
center of Taub-NUT. Before the spectral flow the D5 branes can move
anywhere in Taub-NUT, and this corresponds to motion on the Coulomb
branch. We can now choose a specific Coulomb branch configuration, in
which the D5 branes have moved at a certain distance away from the
Taub-NUT center, and are uniformly distributed on the Taub-NUT fiber,
so as to preserve the Taub-NUT isometry. Hence, from the point of view
of the $\IR^3$ base of the Taub-NUT space, the D5 branes sit at a
point away from the pole of the Taub-NUT harmonic function $V$. This
solution depends on two harmonic functions in $\IR^3$: the D5
harmonic function $Z_2$ is sourced at at the location of the D5
branes, and the Taub-NUT harmonic function $V$ is sourced at the
center of Taub-NUT. After the spectral flow, the new solution can be
straightforwardly obtained from (\ref{uglymetric2}) by setting
$Z_1=Z_3=1$ and $\gamma_2=0$.

The new warp factors and rotation parameter are
\be
\widetilde \mu = - \gamma_1 Z_2 V^{-1}~,~~~\widetilde Z_1 = 1+\gamma_1^2 Z_2 V^{-1}~,~~~ \widetilde Z_2=Z_2~,~~~\widetilde Z_3=1
\ee
and the metric describes a supersymmetric D1-D5 supertube in a Taub-NUT space with nontrivial Wilson lines.
\bea
ds^2_{IIB} &=& \ds \frac{1}{\sqrt{\widetilde Z_1\widetilde Z_2}}\left(- (dt + \widetilde \mu(d \psi + \widetilde A))^2  + (dy - dt  - \gamma_1 \, v_2  )^2 \right) \nonumber \\
&+& \sqrt{\widetilde Z_1 \widetilde Z_2} \left( V^{-1} (d \psi +  A)^2 + V ds^2_{\IR_3}\right)
\label{D5-flow-metric}
\eea
In the new solution both the D5 and the D1 warp factors $ \widetilde Z_2$ and $ \widetilde Z_1$ are sourced at the previous location of the D5 branes, and the Taub-NUT harmonic function $V$ is sourced at the center of Taub-NUT. The solution also has a nontrivial KKM dipole charge with special direction along the common D1-D5 direction and wrapping the Taub-NUT fiber, as well as a nontrivial momentum along the $\psi$ fiber. Note that this solution differs from that of a usual supertube in Taub-NUT \cite{Bena:2005ay}, in that only the D5 harmonic function $L_2$ has a pole at the location of the tube. The D1 harmonic function $L_1$ does not have a pole; the pole in $\widetilde Z_2$ comes rather from the magnetic dipole charge of the supertube interacting with the magnetic flux of the background. Nevertheless, as explained in \cite{Bena:2008dw}\footnote{And as it can be checked  directly from equations (3.36), (3.37) and (3.40) in that paper},  this Taub-NUT supertube is smooth in this duality frame. In fact, the six-dimensional solution has no singular sources.
On the other hand, since the $\psi$ fiber shrinks at the pole of $V$ and the $y$ fiber shrinks at the supertube location, the solution has a topologically nontrivial three-sphere wrapped by $F_3$ flux in the infrared, and the D5 charge of the solution comes now this nontrivial $F_3$ flux.

Hence, what before the generalized spectral flow was a Coulomb branch motion of a D5 brane is now a nucleation of a nontrivial three-sphere. The memory of the D5 brane is only preserved in the RR 3-form flux on this 3-sphere. This is in fact a textbook example of a geometric transition: we started from D5 branes, and we have obtained a the new geometry that has a different topology and the D5 brane charge has been exchanged for flux.

Another configuration in this class would correspond to putting two stacks of D5 branes smeared along the Taub-NUT fiber, at different points in the $\IR^3$ base of Taub-NUT. These would give rise to two KK monopoles, each one at the position of the D5's.  There will be two 3-spheres between these KKM's and the origin, and the D5 charge will come from $F_3$ flux on these three-spheres.

It is also very likely that this round supertube in Taub-NUT will represent just a particular configuration on the moduli space of the new theory. As one can find both from their DBI description \cite{Mateos:2001qs} and from their supergravity solution \cite{Bena:2008dw}, supertubes can have arbitrary shape, and their classical moduli space is parameterized by several continuous functions, and its dimension is therefore infinity. It would be interesting to perform a DBI analysis of the supertube we have found to check whether this expectation is indeed confirmed or whether, due to the absence of a pole in the D1 harmonic function $L_1$, this supertube is in fact rigid.

\subsubsection*{The effect of a the second Melvin twist on D1 branes}

Let us consider now a solution describing D1 branes on a particular Coulomb branch configuration where they are smeared on the Taub-NUT circle at a certain location in the Taub-NUT base, and perform the $ \lambda_2$ transformation which corresponds to 5 T-dualities, a Melvin twist and 5 T-dualities back, or a generalized spectral flow with a coefficient $\gamma_2$. The original solution has a pole in $V $ at the center of Taub-NUT, and a pole in $Z_1$ at the D1 location in the $\IR^3 $ base of Taub-NUT, while $Z_2 = Z_3 = 1$ and $\gamma_1 = 0$. The resulting solution is again given by (\ref{D5-flow-metric}), except that now
\be
\widetilde \mu = - \gamma_2 Z_1 V^{-1}~,~~~\widetilde Z_2 = 1+\gamma_2^2 Z_1 V^{-1}~,~~~ \widetilde Z_1=Z_1~,~~~\widetilde Z_3=1
\ee
This solution describes again a smooth D1-D5 supertube in Taub-NUT, and this supertube again differs from the usual one in that now the D5 harmonic function, $L_2$, has no pole at the location of the tube, and the divergence in the D5 warp factor $\widetilde Z_2$ comes from the interaction of the magnetic dipole charge of the supertube with the magnetic field of the background. This is again a smooth supersymmetric solution, as can be checked either directly or using equations (3.36), (3.37) and (3.40) in  \cite{Bena:2008dw}.
The difference with respect to the previous solution is that now the geometric transition has exchanged the singular D1 brane charges for a nontrivial flux of $*_6 F_3$ on the $S^3$ in the infrared (or from a ten-dimensional perspective with a nontrivial flux of $F_7$ on $S^3 \times T^4$). Again, we expect a DBI analysis of these supertubes to reveal a much larger moduli space.

\subsubsection*{Two generalized spectral flows on D1 and D5 branes}

Given that we now have a powerful way of following the solutions through the Melvin and spectral flow transformations, one can also ask what happens to the Coulomb branch of the original D1-D5 system after the two generalized spectral flows that take us to geometries with a NHEK infrared.

One can consider first a circular Coulomb branch distribution, in which the D1's and D5's are smeared on the Taub-NUT fiber at different positions in the $\IR^3$ base. The new solution will have KKM dipole charges along the common D1-D5 direction at the location of the D1 and the D5 branes. Moreover, since the original stacks of D1 and D5 branes were both locally preserving 16 supersymmetries, the resulting near-brane solution will also preserve 16 supersymmetries. Furthermore, after the two spectral flows, the centers will acquire a nontrivial KKM charge with a $\psi$ special direction, and hence will give rise to a smooth bubbling solution \cite{Bena:2008wt}, of the type constructed in \cite{Bena:2005va,Berglund:2005vb,Saxena:2005uk} to describe black hole microstate solutions\footnote{If one dualizes this solution to the frame where the D1, D5 and P charges correspond to three sets of M2 branes on $T^2$s inside $T^6$, and then reduces this solution to ten dimensions along $\psi$, each smooth center will become a fluxed D6 brane \cite{Balasubramanian:2006gi,Bates:2003vx}. However, unlike the usual bubbling solutions, the positions of our bubbles are not constrained by any bubble equations.}.

In our IIB frame this bubbling solution will have nontrivial three-cycles wrapped by flux. These three-cycles can be thought of as coming from the geometric transition of the D1 branes and the D5 branes. The three-cycles between the center of Taub-NUT and the D5 locations will have a nontrivial $F_3$ flux, equal to the numbers of D5 branes before the geometric transition.  The three-cycles between the center of Taub-NUT and the D1 locations will have a nontrivial flux of $*_6F_3$, equal to the numbers of D1 branes before the transition.

Hence, at generic points on this bubbling branch of the moduli space, where the D1 and the D5 shells do not coincide, the overall solution will be smooth. When the two shells coincide,  the resulting center on the Taub-NUT base will locally have  eight supersymmetries, and despite the presence of KKM charges will not be smooth. Much like the configurations in the previous subsections, we expect these bubbling geometries to represent but a small subset of the configurations of the new moduli space. Indeed, from the physics of supertubes it is likely that there should exist supersymmetric deformations of bubbling geometries parameterized by several continuous functions. It would be clearly interesting to construct them.

\subsubsection*{More general configurations in the new moduli space}

For a more general configuration, the resulting charges and Coulomb branch configurations will be more complicated. After performing two spectral flows, the resulting configuration will generically have  all the possible charges and dipole charges it can carry. If the solution near one of the centers had 16 supersymmetries before the spectral flow, the resulting solution will be generically smooth. However, centers that have more than one type of branes before the spectral flows (and hence will only preserve eight or four supersymmetries) will generically not be smooth after the flows. 

It is also important to note that the branch of the moduli space we discussed above will always fit into the NHEK region of the solution, and will therefore correspond to deformations of the asymptotically NHEK solution. This is because  the distance between the center of Taub-NUT and the branes was not constrained in the Coulomb branch configurations we started from (in the language of multicenter solutions the charges at all centers are mutually local, and hence there are no bubble/integrability equations to constrain their positions). Hence, after the generalized spectral flows one still does not have any integrability conditions to satisfy, and the size of the bubbles will again be a free parameter.

Note that this will not necessarily be true for the Higgs branch configurations. Before the generalized spectral flows some of these configurations are supertubes, whose distance from the Taub-NUT center are constrained. Hence, they may or may not survive when taking the near-horizon limit that brings us to the NHEK geometries. We leave the exhaustive exploration of these configurations to a future publication.

\subsection*{Acknowledgements}

We are grateful to Sheer El-Showk, Stefano Giusto, Kyriakos Papadodimas, Clement Ruef and Andrew Strominger for interesting conversations, and to Andrew Strominger in particular for useful comments on the draft. M.G. would like to thank the IPhT, Saclay and the Center for the Fundamental Laws of Nature at Harvard for their kind hospitality.
The work of I.B. is supported in part by the ANR grant 08-JCJC-0001-0, and by the ERC Starting Independent Researcher Grant 240210 - String-QCD-BH.  The work of M.G. is supported in part by the DOE grant DE-FG02-95ER40893, and the work of W.S.  by the Harvard Society of Fellows and the DOE grant DE-FG02-91ER40654. I.B. and M.G. are also grateful to the  Aspen Center for Physics for hospitality and 
support via the NSF grant 1066293. 

\appendix

\section{Hodge duals \label{hodge}}

We present the Hodge duals of various three-forms of interest with respect to the metric \eqref{bck}. Our conventions are such that $\e_{\tau \rho y\th\phi\psi} =1$.

\bea
\star(\s_1 \wedge \s_2 \wedge \s_3) & = & - \frac{1}{2 \sqrt{1-\e_g^2}}  w_+ \wedge w_- \wedge (w_3 + \e_g \s_3) \non \\
\star(w_+ \wedge w_- \wedge w_3) & = & - \frac{2}{ \sqrt{1-\e_g^2}} \, \s_1 \wedge \s_2 \wedge (\s_3 + \e_g w_3) \non \\
\star(\s_1 \wedge \s_2 \wedge w_3) & = & \frac{1}{2 \sqrt{1-\e_g^2}}  w_+ \wedge w_- \wedge (\s_3 + \e_g w_3) \non \\
\star(w_+ \wedge w_- \wedge \s_3) & = & \frac{2}{ \sqrt{1-\e_g^2}} \, \s_1 \wedge\s_2 \wedge (w_3 + \e_g \s_3) \, .
\eea
Therefore, the Hodge dual of the three-form field strength for the $\varepsilon = - 1$ backgrounds is

\bea
\star H & = &  \frac{\ell^2}{\sqrt{1-\e_g^2}} \left[ (1 + \frac{ \g \e_B \e_g}{h}) \left( \s_1  \wedge \s_2 \wedge \s_3 + \half w_+ \wedge
w_- \wedge w_3\right) + \right. \non \\
&& \left. \hspace{3 cm} + ( \e_g + \frac{ \g \e_B}{h}) \left( \s_1  \wedge \s_2 \wedge w_3 + \half w_+ \wedge
w_- \wedge \s_3\right)  \right]
\eea
The above three-form field is no longer of the form \eqref{bck}: it differs by an overall sign. Nevertheless, the interchange of $\phi$ and $\psi$ which we are supposed to perform for the $\varepsilon =-1$ backgrounds effectively reverses the sign of the Levi-Civita tensor, thus yielding a three-form field of the expected form.

\section{$SO(2,2)$ transformations \label{so22}}

In general, under an $SO(d,d)$ T-duality transformation, the internal metric, B-field and the Kaluza-Klein gauge fields transform as \cite{schw}

\be
 X \r (\Om_{22} X + \Om_{21})(\Om_{11} + \Om_{12} X)^{-1} \;, \;\;\;\;\; A^i \r \Om^i{}_j A^j
\ee
where $A^i = \{\A^\a,\B_\a \}$ has been defined below \eqref{kkred} and we let

\be
X \equiv  G+B
\ee
The $2d \times 2d$ dimensional matrix $\Om$ satisfies

\be
\Om^T \eta\, \Om =\eta\;, \;\;\;\;\;\eta = \left(\begin{array}{cc} \;0\; & \;\,\mathbb{I}_d\; \\ \, \mathbb{I}_d & 0 \end{array}\right)
\ee
The $SO(2,2)$ transformations are particularly  simple \cite{porra}. Parameterizing

\be
X = \frac{\rho_2}{\tau_2} \left(\begin{array}{cc} |\tau|^2 & \tau_1 \\ \tau_1 & 1 \end{array}\right) +
\rho_1  \left(\begin{array}{cc} 0 & \;\;1\;\; \\ -1 & 0 \end{array}\right)
\ee
we immediately see that complex structure transformations of the underlying torus

\be
\tau \r \frac{a \tau +b}{c \tau + d}
\ee
act on $X$ simply as

\be
X \r \Lambda X \Lambda^T \;, \;\;\;\;\; \Lambda =  \left(\begin{array}{cc} a & b \\ c & d \end{array}\right)
 \;\;\;\; \Rightarrow \;\; \Om =  \left(\begin{array}{cc} (\Lambda^T)^{-1} & 0 \\ 0 & \Lambda \end{array}\right)
\ee
Meanwhile, the K\"{a}hler structure transformations

\be
\rho \r \frac{a \rho +b}{c \rho + d} \label{khlrho}
\ee
are implemented by

\be
\Om =  \left(\begin{array}{cc} d & - c \,\e \\ b \,\e & a \end{array}\right)\;, \;\;\;\;\;\e= \left(\begin{array}{cc} \;\;0\; & \;1 \;\\ -1 & 0 \end{array}\right) \, .\label{kaehlert}
\ee

\bigskip

\bigskip

\noindent Let us now discuss the structure of the $SL(2,\mathbb{R})_L \times SU(2)$ invariant backgrounds that we would like to study. There are two coordinate systems that we can consider. The solutions \eqref{bck}
yield a metric $\hat g_{mn}$ and B-field that have a relatively simple structure. Nevertheless, the $SO(2,2)$ transformations naturally act on the metric written in the $y^\a$ coordinate system, where all compact coordinates have period $2 \pi$. Given the coordinate transformation \eqref{defm} between the two, the quantities of interest are related by

\be
G = M^T \hat G M \;, \;\;\; B = M^T \hat B \,M =\hat  B\, \det M \;, \;\;\;\;\; \A =
M^{-1} \hat \A \;, \;\;\; \B = M^T \hat \B
\ee
The internal metric $(\hat G)$, B-field $(\hat B)$ and Kaluza-Klein gauge fields $\hat \A, \hat \B$ in the $\hat y^\a$ coordinate system, evaluated on the backgrounds \eqref{bck}, take a particularly simple form

\be
\hat \A^{ y} = r dt \;, \;\;\;\;\; \hat \A^{ \psi} = \cos \th d \phi \label{aref}
\ee
The gauge fields which descend from the Kaluza-Klein reduction of the $6d$ B-field are defined as \cite{schw}

\be
\hat \B_{\mu\a} = \hat B_{\mu\a} + \hat B_{\a\b}\, \hat \A_\mu{}^\b
\ee
where $\hat B$ is the six-dimensional gauge field in the respective coordinate system, given by

\be
\hat B = \ell^2  \bigl( {-} r dt \wedge d y - \varepsilon \cos\th d\phi \wedge d\psi \bigr) + \frac{\ell^2 \g \e_B}{ h} \, w_3 \wedge \s_3 +
\Delta B \, dy \wedge d\psi
\ee
The $\Delta B$ term is a constant pure gauge term which we had to include, as it gives a nonzero contribution to $\hat \B$

\be
\hat \B_y =  {-}\ell^2  \, r dt + \Delta B \, \cos\th d\phi \;, \;\;\;\;\; \hat \B_\psi = - \varepsilon \,\ell^2\cos\th d\phi -
\Delta B \, r dt
\ee
Put more simply, if we consider $\hat \A^\a$ in \eqref{aref} to be a reference gauge field for these backgrounds, due to its simple form,  then

\be
\hat \B = ( {-}\ell^2 \,I_\varepsilon + \Delta B \, \e ) \hat \A  \;, \;\;\;\;\;\;\; I_\varepsilon \equiv \left(\begin{array}{cc}  1 & \;\;\;0\; \\ 0\; &  \varepsilon \end{array}\right)
\ee
In the $\hat y^\a$ coordinate system, the internal metric and B-field are given by 

\be
\hat G_{\a\b} = \frac{\ell^2 \g}{h} \left(\begin{array}{cc}  1 & \;\e_g\; \\ \;\e_g\; & 1\end{array}\right) \;, \;\;\;\;
\hat B_{\a\b} = \left(\frac{\ell^2 \g \e_B}{h} + \Delta B\right) \left(\begin{array}{cc} \; 0 & \;1\; \\ -1\; & 0\end{array}\right)
\ee

\subsection{K\"{a}hler transformations}

We start from a spacetime whose curvature is characterized by $\e_g, \e_B$ and $\ell$, and global structure given by the identification matrix $M$, of determinant $\det M  = \ell^{-2} \V$. We will assume that the starting $\Delta B =0$, which can be always achieved by a constant gauge transformation. The starting Kaluza-Klein gauge fields are thus

\be
\A = M^{-1} \hat \A \;, \;\;\;\;\; \B =  {-} M^T \ell^2 \, I_\varepsilon \hat \A
\ee
After acting with a K\"{a}hler transformation of the type \eqref{kaehlert}, the new Kaluza-Klein gauge fields must take the form

\be
\A' = M'^{-1} \hat \A \;, \;\;\;\;\; \B' = M'^{T} ( {-}\ell'^2 \, I_\varepsilon + \Delta B' \, \e )\, \hat \A
\ee
These equations  determine the new identification matrix $M'$, which is given by

\be
M' = \frac{d  {-} c \V \, \e \,I_\varepsilon}{d^2  {+} c^2 \V^2 \varepsilon} \, M  \;, \;\;\;\;\; \V = \ell^2 \det M \label{solmp}
\ee
and the parameters $\ell'$ and $\Delta B'$
\be
\ell' = \ell\;, \;\;\;\;\; \Delta B' = \ell^2 \left( \frac{b\, d }{\V}  {+} a \, c \,\varepsilon\, \V  \right)
 \label{soldb}
\ee
Let us moreover assume that the original background has $\e_B =0$. Then the internal B-field $\rho_1 =0$ and

\be
\g = 1 \;, \;\;\;\;\; h = \sqrt{1-\e_g^2}\;, \;\;\;\;\; \rho_2= \sqrt{\det G} = \ell^2 \det M = \V
\ee
Acting with the K\"{a}hler structure transformation \eqref{khlrho}, we find
\be
\rho'_1 = \frac{b d + a c \V^2}{d^2 + c^2 \V^2} \;, \;\;\;\;\; \rho'_2 = \frac{\V}{d^2 + c^2 \V^2}
\ee
From the explicit expression for $\rho'_1$

\be
\rho'_1 = \left(\frac{\ell'^2 \g' \e'_B}{h'} + \Delta B'\right) \det M'
\ee
and the solution for $\Delta B'$ and $M'$, we deduce that

\be
\frac{\e_B'}{\sqrt{1-{\e'_g}^2}} = \frac{c d \V (1- \varepsilon)}{d^2 -  c^2 \V^2} \;, \;\;\mbox{for}\;\;\; |d| > |c|\V \label{eqeB}
\ee
The extra equation we need in order to determine the geometry is obtained from the fact that K\"{a}hler transformations do not change the complex structure of the compactification torus
\be
\M_\tau = \frac{1}{\tau_2} \left(\begin{array}{cc} |\tau|^2 & \tau_1 \\ \tau_1 & 1\end{array} \right)=\frac{1}{\sqrt{\det G}} \, G_{\a\b} = \frac{1}{\det M } \, M^T \hat \M \, M
\ee
where $\hat \M$ is defined as

\be
\hat \M = \frac{1}{\sqrt{\det\hat G}} \, \hat G_{\a\b} = \frac{1}{\sqrt{1-\e_g^2}} \left( \begin{array}{cc} 1 & \;\e_g \\ \;\e_g & 1 \end{array} \right)
\ee
Consequently, we have

\be
\M_\tau = \frac{1}{\det M }\, M^T \hat \M \, M=
\frac{1}{\det M' }\, M'^T \hat \M' \, M'
\ee
with solution

\be
\hat \M' = \frac{1}{d^2  {+} c^2 \V^2 \varepsilon} (d- c \V I_\varepsilon \e) \, \hat \M \, (d+ \e I_\varepsilon c \V) \label{eqnm1}
\ee
This equation immediately allows us to determine $\e_g'$ in terms of $\e_g$.

\subsubsection*{Summary}

In practice, what we need to know are the parameters $\e'_g, \e'_B, \ell', M', \Delta B'$ ad $\Phi'$ of the new background that results from performing a K\"{a}hler transformation upon a background characterized by $\e_g, \ell, M, \Phi$ and $\e_B = \Delta B =0$.

The answer is that $\e'_g$ is determined by $\e_g$ via \eqref{eqnm1}. For $\varepsilon =-1$, one finds that

\be
\e'_g = \frac{ \e_g -x }{1 - x \,\e_g} \;,\;\;\;\;\; x = \frac{2 c d \V}{d^2 + c^2 \V^2}
\ee
For $\varepsilon =1$ we simply obtain the consistency check that if we start with $\e_g=0$, or $\hat \M=\mathbb{I}_2$, we also end up with $\hat \M' = \mathbb{I}_2$, or $\e'_g=0$. For $\varepsilon =-1$  the expression for $\e'_B$ is determined by \eqref{eqeB} and the above expression for $\e'_g$ to be

\be
\e'_B = \frac{x \sqrt{1-\e_g^2}}{1- x \, \e_g} 
\ee
The remaining parameters are given in \eqref{solmp}, \eqref{soldb}. The value of the dilaton for the new background is computed from the requirement that the four-dimensional dilaton $e^{-2 \phi_4} = e^{-2 \Phi} \sqrt{\det G}$ be unchanged by the K\"{a}hler transformation, yielding

\be
e^{2 \Phi'} = \frac{1}{d^2 + c^2 \V^2} \, e^{2 \Phi}
\ee

\section{Details of the matching \label{match}}

We first match the parameters of the $\mathcal{T}$ and $\mathcal{S}$ transformations by requiring that the final $\e_{g,B}$ be given by \eqref{ekerr}.
From the formulae in  section \ref{stu}, we immediately deduce that
\be
x=  - \frac{1}{\cosh 2 \d_1} \;, \;\;\;\;\; \tilde{x} = -\frac{1}{\cosh 2 \d_5}
\ee
from which we find that the parameters of the two transformations are

\be
 \frac{c \V}{d}= - e^{\pm 2\d_1}  \;, \;\;\;\;\;   \frac{\tilde{c} \tilde{\V}}{\tilde{d}} = - e^{\pm 2 \d_5}
\ee
Given that in the limit $\d_i \r \infty, a \r 0$ with the charges kept fixed we are supposed to recover $AdS_3 \times S^3$ automatically, we choose the lower signs. After performing an S-duality to the NS1-NS5 system we have

\be
e^{2\Phi'_2} = \frac{\cosh 2 \d_5}{\cosh 2 \d_1} \;, \;\;\;\;\; \tilde{L}^{'2} = \frac{a^2}{2} \sinh 2 \d_5
\ee
From here, we obtain the following relations

\be\,
\ell \tilde{d} = \frac{a e^{\d_5}}{2} \;, \;\;\;\;\; \frac{\tilde{d} e^{\Phi_0}}{d} = e^{\d_5-\d_1} \;, \;\;\;\;\;  \tilde{c}\, \tilde{d} =c \, d
\ee
where the last relation has been derived using \eqref{relvol}. Noting that in the original NS-NS background

\be
\ell^2 = \frac{q_5}{4} \;, \;\;\;\;\; e^{2\Phi_0} = \frac{q_5}{q_1}
\ee
we can use the above relations to express the parameters $Q_1, Q_5, Q_p$ and $J_L$ of the resulting Kerr black hole in terms of  $q_1, q_5, q_0, c, d, \tilde{c}, \tilde{d}$. We find that

\be
e^{-2 \d_1} = - \frac{c \hat S_0}{d q_1} \;, \;\;\;\;\; e^{-2 \d_5} = - \frac{\tilde{c} \hat S_0}{\tilde{d} q_5}\;, \;\;\;\;\; a^2 = - c d \hat S_0 \label{pbh}
\ee
where we have defined the ``reduced'' entropy

\be
\hat S_0 \equiv \sqrt{q_1 q_5 q_0}
\ee
This implies that the integer charges $Q_1$ and $Q_5$ are

\be
Q_1 = q_1 d^2 - q_5 q_0 c^2 \;, \;\;\;\;\;
Q_5  = q_5 \tilde d^2 - q_1 q_0 \, \tilde c^2 \label{solQ}
\ee
Thus, if the input charges and the coefficients of the transformations are integer, then the D1 and D5 charges of the resulting NHEK background are also automatically integer. Let us now proceed to matching the identifications. We have

\be
M_2 = \frac{1}{(d^2 - c^2 \V^2)(\tilde{d}^2-\tilde{c}^2 \tilde{\V}^2)} \left(\begin{array}{cc} \tilde{d} & \;\;\tilde{c}\tilde{\V} \\  \tilde{c} \tilde{\V} & \tilde{d}\end{array}\right) \left(\begin{array}{cc} d & \;\; c\V \\  c\V & d\end{array}\right) M_0
\ee
In order for the match to happen, we need

\be
M_2 \, \mathcal{U} = M_{NHEK} = \left(\begin{array}{cc} 2 \pi T_Q & \;\;0\;\; \\ - \frac{2J_L}{Q_1 Q_5} & 2\end{array}\right) \label{tomatch}
\ee
where $\mathcal{U}$ is the unimodular matrix \eqref{stumat} representing the final $\mathcal{U}$ transformation.  First, note that

\be
\det M_2 = \frac{\det M_0}{(d^2 - c^2 \V^2)(\tilde{d}^2-\tilde{c}^2 \tilde{\V}^2)} = 4 \pi T_Q = \frac{2 S}{\pi Q_1 Q_5} \label{deteqn}
\ee
where $S$ is the entropy of the $5d$ Kerr black hole, given in \eqref{entkerr}. Manipulating the above equation we find, as expected, that the entropy of the black hole is invariant under the series of STU dualities

\be
\hat{S}_0 = \frac{S}{2\pi}= \sqrt{ J_L^2 - Q_1 Q_5 Q_p} =  \sqrt{q_1 q_5 q_0} \label{entro}
\ee
This fact lets us determine the remaining charges, $Q_p$ and $J_L$, in terms of $q_i, c, \tilde c, d, \tilde d$. Remember that
 the expression for the entropy $\hat S_0$ in terms of the $\d_i$ is

\be
\hat S_0 = 4 a^3 (c_1 c_5 c_p - s_1 s_5 s_p) \label{sofd}
\ee
which can be viewed as a quadratic equation for $e^{\d_p}$

\be
2 a^3 \cosh (\d_1 -\d_5) \, e^{2\d_p} - \hat S_0\, e^{\d_p} + 2 a^3 \cosh (\d_1 + \d_5) =0
\ee
Using \eqref{pbh}, one can simplify the discriminant of this equation to

\be
\Delta = \hat S_0^2 \left[ 1 - 4 c^2 (d^2 q_1 + \tilde d^2 q_5) (d^2 + \tilde{c}^2 q_0)\right]
\ee
In order to have real solutions for $e^{\d_p}$, the discriminant must be positive; this requirement cannot be satisfied for integer $q_i, c, \tilde{c}, d, \tilde d$. Consequently, there does not exist a string theory duality which maps $AdS_3 \times S^3$ to NHEK, but only a supergravity one. This being said, one can solve the above  equation for $\d_p$ and then calculate $Q_p$ and $J_L$, using

\be
Q_p = 2 a^2 \sinh 2 \d_p \;, \;\;\;\;\; J_L = 4 a^3 (c_1 c_5 c_p + s_1 s_5 s_p)
\ee
The simplest way to write the resulting formulae seems to be

\be
J_L = \frac{\cosh^2 (\d_1 + \d_5) + \cosh^2 (\d_1 - \d_5)}{2 \cosh (\d_1 + \d_5) \cosh (\d_1 - \d_5)} \, \hat S_0 +  \frac{\cosh^2 (\d_1 + \d_5) - \cosh^2 (\d_1 - \d_5)}{2 \cosh (\d_1 + \d_5) \cosh (\d_1 - \d_5)} \,\sqrt{\Delta}
\ee

\be
Q_p = \frac{1}{16 a^4} \left( \frac{(\hat S_0 + \sqrt{\Delta})^2}{\cosh^2(\d_1 - \d_5)} - \frac{(\hat S_0 - \sqrt{\Delta})^2}{\cosh^2(\d_1 + \d_5)}  \right)
\ee
We are finally ready to study the identifications. Letting

\be
M_0 = \left(\begin{array}{cc} 2 \pi T_0 & 0 \\ 0 & 2 \end{array} \right)
\ee
and

\be
\mathcal{U} = \left(\begin{array}{cc}a' & b' \\ c' & d' \end{array} \right) \;, \;\;\;\;\; a' d' - b' c' =1
\ee
we find that 

\be
a' = \frac{1 + e^{2\d_1+2\d_5} - \frac{e^{2\d_1} + e^{2\d_5}}{\pi T_R}}{4  d \tilde{d} \sinh 2 \d_1 \sinh 2 \d_5} \;, \;\;\;\;\; c' = \pi d \tilde{d} T_Q (e^{-2 \d_1} + e^{-2 \d_5} - \frac{1+ e^{-2 \d_1 - 2 \d_5}}{\pi T_R})
\ee

\be
d' = d \tilde{d}  (1+ e^{-2\d_1-2\d_5})\;, \;\;\;\;\; b' = \frac{d \tilde d}{\pi T_0} (e^{-2\d_1} + e^{-2\d_5}) \label{finu}
\ee
\iffalse
In terms of the integer charges, the parameters  are  {\emph{Out?}}

\be
d' = \frac{\tilde d }{q_1 d} (q_1 d^2 + q_5 \tilde{d}^2 - Q_5) \;, \;\;\;\;\; b' = \frac{c}{\tilde d} (q_1 d^2 + q_5 \tilde d^2)
\ee
etc.
\fi
Thus, the final charges $Q_1, Q_5, Q_p, J_L$ are entirely determined by the input charges $q_1, q_5, q_0$ and the transformation parameters $c, \tilde c, d, \tilde d$ via the formulae written above. The parameters of the transformations cannot all be chosen to be integers.

Nevertheless, what we would normally be more interested in is the inverse problem: given a NHEK near-horizon characterized by the charges $Q_1, Q_5, Q_p$ and $J_L$, can one find  a set of STU transformations which un-twist it to an  $AdS_3 \times S^3$ geometry, and how many of the  parameters $q_1, q_5, q_0, c, \tilde c, d, \tilde d$ are fixed by the final data $Q_i, J_L$ and how many are free. A simple counting shows that four of the initial parameters are determined and another two are free, but in order to see explicitly which ones are fixed it is useful to work out an example.

\subsubsection*{Example: Equal charges}

We would like to find the STU transformations needed to untwist the near horizon geometry of an extremal Kerr black hole with $Q_1=Q_5=Q_p =Q$ and angular momentum $J_L$. We will only work out the $\mathcal{S}$ and $\mathcal{T}$ transformations, as they are the most nontrivial ones in going from $AdS_3 \times S^3$ to $6d$ NHEK.
Setting $\d_1=\d_5$ and using \eqref{pbh} we find that

\be
q_1 d^2 = q_5 \tilde d^2
\ee
We can think of this equation as determining $q_5$ in terms of $q_1, d$ and $\tilde d$. Next, plugging in the solution \eqref{pbh} for $\d_1$ and $a$ into the equation \eqref{sofd} with $\d_p = \d_5 =\d_1$ we find that

\be
 c \sqrt{q_1} \left(3 d^2 + \frac{c^2 q_5 q_0}{q_1} \right) = -1 \label{conse}
\ee
The equation \eqref{solQ} implies that

\be
c^2 = \frac{q_1 d^2 - Q}{q_5 q_0} \label{formc}
\ee
which can be used to manipulate \eqref{conse} into a cubic equation for $q_1 d^2$

\be
 (4 q_1 d^2 - Q)^2 (q_1 d^2-Q) = \hat S_0^2
\ee
where now $\hat S_0$ should be thought of as a function of $Q$ and $J_L$, namely \eqref{entro}. The above equation has one real solution with $q_1 d^2 >Q$. Once this solution is known, one can determine the combinations

\be
c\, d = - \frac{\sqrt{(q_1 d^2 -Q) q_1 d^2}}{\hat S_0} = \tilde c\, \tilde d
\ee
Using the fact that the solution for $q_1 d^2>Q >1$, one immediately notes that $|c \, d| <1$, so again we confirm that the transformations' parameters cannot all be integer. The remaining charges are determined as

\be
q_5 = \frac{q_1 d^2}{\tilde d^2} \;, \;\;\;\;\; q_0 = \frac{\hat S_0^2}{q_1 q_5} = \frac{ \tilde d^2 \hat S_0^2 }{q_1^2 d^2}
\ee
Thus, the charges $Q, J_L$ of the Kerr-Newman black hole we aim for determine the combinations  $q_1 d^2, q_5\, \tilde{d^2}, q_0 \,d^{-2} \tilde{d}^{-2}, c\,d, \tilde c\, \tilde d $, but do not seem to impose any constraint on $d, \tilde d$. A natural choice is then to take $d = \tilde d =1$. Since the only constraint on $a,\tilde a, b, \tilde b$ is the unimodularity one, we can also choose for simplicity $a = \tilde a =1$, $b = \tilde b =0$, and thus the $\mathcal{S}$ and $\mathcal{T}$ transformations become TsT's.

\section{An identity among dualities \label{dualid} }

In section \ref{spflgeom} we showed how the metric and three-form flux transform under the first generalized spectral flow, or S\,TsT\,S. Now we will show that a T-duality, followed by a Melvin twist and a T-duality back in a purely RR background has the same effect on a given geometry.

After a T duality along $y^2$, the background (\ref{metfib}) becomes

\be
ds^2_{10}=ds^2_8+G_{22}^{-1}(dy^{2})^2+\det{G} \, G_{22}^{-1}(dy^{1}+\mathcal{A}^{1})^2
\ee

\be
B=\left(\mathcal{A}^{2}+{G_{12}\over G_{22}}(dy^{1}+\mathcal{A}^1)\right)\wedge dy^2 \;, \;\;\;\;\;
e^{2\phi'}=G_{22}^{-1}e^{2\phi}
\ee

\be
C_{(1)}={C}_{\mu2}dx^\mu+\zeta dy^1 \;, \;\;\;\;\;\;
C_{(3)}=-\left(\half C^{(2)}_{\mu\nu}dx^\mu\wedge dx^\nu+C^{(2)}_{\mu1}\wedge dy^{1}\right)\wedge dy^2 \non
\ee
Next, we lift to M-theory, using

\be
 ds^2_{11}=e^{-2\phi'\over3}ds_{10}^2+e^{4\phi'\over3}(dx^{11}+C_{(1)})^2
\ee
Shifting $y^1\rightarrow y^1+\hat{\lambda}_1 x^{11}$ and reducing to ten dimensions, we obtain

\be
ds^2_{10A}=\Sigma_1^{1\over2}(ds_8^2+G_{22}^{-1}(dy^{2})^2)+\Sigma_1^{-{1\over2}}\det{G}\,G^{-1}_{22}(dy^1+\mathcal{A}^{1'})^2
\ee

\be
B'=\left(\mathcal{A}^{2'}+{G_{12}\over G_{22}}(dy^1+\mathcal{A}^{1'})\right)\wedge dy^{2}\;,\;\;\;\;\;
e^{2\phi''}={G_{22}^{-1} \Sigma_1^{3\over2}}e^{2\phi}
\ee

\be
C'_{(3)}=-\left(\half \mathcal{C}^{(2)}_{\mu\nu}dx^\mu\wedge dx^\nu+\mathcal{B}_1\wedge(dy^1+\mathcal{A}^{1'})\right)\wedge dy^1\;,\;\;\;\;\;
C'_{(1)}=\mathcal{A}^{1A}
\ee
where
\be
\Sigma_1=(1+\hat{\lambda}_1\zeta)^2+\hat{\lambda}_1^2e^{-2\phi}\det{G}\;, \;\;\;\;\;
\mathcal{A}^{'\alpha}=\mathcal{A}^\alpha-\hat{\lambda}_1\hat{\epsilon}^{\alpha\beta}\mathcal{B}_\beta
\ee
\bea
\mathcal{A}^{1A}&=&\Sigma_1^{-1}\left((1+\hat{\lambda}_1 \zeta)(\mathcal{B}_2+\zeta(dy^1+\mathcal{A}^1))+\hat{\lambda}_1 \det{G} e^{-2\phi}(dy^1+\mathcal{A}^1) \right)\nonumber\\&=&\mathcal{B}_2+\zeta'''(dy^1+\mathcal{A}^{1'})\non\eea
where $\zeta'''$ is given in \eqref{newparamone}. A final T-duality along $y^1$ gives

\be
ds^{2'''}_{10}=\Sigma_1^{1\over2}ds_8^2+\Sigma_1^{-{1\over2}} G_{\alpha\beta}(dy^{\alpha}+\mathcal{A}^{\alpha'})(dy^\beta+\mathcal{A}^{\beta2'})
\ee

\be
B=0\;, \;\;\;\;\;
e^{2\phi'''}=e^{2\phi}\Sigma_1
\ee
\be
C'''_{(2)}=\half\, \mathcal{C}^{(2)}_{\mu\nu}dx^\mu\wedge dx^\nu+\mathcal{B}_\alpha\wedge(dy^\alpha+\mathcal{A}^{\alpha'})+\zeta'''(dy^2+\mathcal{A}^{2'})\wedge (dy^1+\mathcal{A}^{1'})
\ee
The above solution is the same as the solution after STsTS given by (\ref{tstmet}). Therefore, we have explicitly shown that
 STsTS on a IIB background with only RR three-form flux is equivalent to a T-duality, Melvin twist, and another T-duality.

%%%%%%%

\end{document}